\documentclass[useAMS,usenatbib,referee]{biom}

\usepackage{subcaption}
\usepackage{adjustbox} 
\usepackage{amssymb}
\usepackage{amsmath}
\usepackage{hyperref}
\usepackage{graphicx}
\usepackage{xcolor}

\usepackage{booktabs}

\def\bSig\mathbf{\Sigma}
\newcommand{\Ebb}{\mathbb{E}}
\newcommand{\Ibb}{\mathbb{I}}
\newcommand{\Pbb}{\mathbb{P}}

\newcommand{\aug}{\text{aug}}

\title{Nonparametric estimation of the total treatment effect with multiple outcomes in the presence of terminal events}{}

\author{Jessica Gronsbell$^{1, *}$ %\emailx{j.gronsbell@utoronto.ca}, 
Zachary R.\ McCaw$^{2,*}$, Isabelle-Emmanuella Nogues$^{2,*}$, \\
{\bf{ Xiangshan Kong$^{1}$, Tianxi Cai$^{2}$, Lu Tian$^{3}$, LJ Wei$^{2}$}} \\
$^{1}$ Department of Statistical Sciences, University of Toronto, Toronto, ON, Canada \\
$^{2}$Department of Biostatistics, Harvard T.H Chan School of Public Health, Boston, MA, U.S.A \\
$^{3}$Department of Biomedical Data Science, Stanford University, Stanford, CA, U.S.A \\
$^{*}$These authors contributed equally. }

\begin{document}

%  This will produce the submission and review information that appears
%  right after the reference section. Of course, it will be unknown when
%  you submit your paper, so you can either leave this out or put in 
%  sample dates (these will have no effect on the fate of your paper in the
%  review process!)

\date{{\it Received December} 2024. {\it Revised XX}. {\it
Accepted XX}.}

%  These options will count the number of pages and provide volume
%  and date information in the upper left hand corner of the top of the 
%  first page as in published papers. The \pagerange command will only
%  work if you place the command \label{firstpage} near the beginning
%  of the document and \label{lastpage} at the end of the document, as we
%  have done in this template.

%  Again, putting a volume number and date is for your own amusement and
%  has no bearing on what actually happens to your paper!  

\pagerange{\pageref{firstpage}--\pageref{lastpage}} 
\volume{}
\pubyear{2024}
\artmonth{}

%  The \doi command is where the DOI for your paper would be placed should it
%  be published. Again, if you make one up and stick it here, it means 
%  nothing!

\doi{}

%  This label and the label ``lastpage'' are used by the \pagerange
%  command above to give the page range for the article. You may have 
%  to process the document twice to get this to match up with what you 
%  expect. When using the referee option, this will not count the pages
%  with tables and figures. 

\label{firstpage}

%  put the summary for your paper here
\begin{abstract}
As standards of care advance, patients are living longer and once-fatal diseases are becoming manageable. Clinical trials increasingly focus on reducing disease burden, which can be quantified by the timing and occurrence of multiple non-fatal clinical events. Most existing methods for the analysis of multiple event-time data require stringent modeling assumptions that can be difficult to verify empirically, leading to treatment efficacy estimates that forego interpretability when the underlying assumptions are not met. Moreover, many methods do not appropriately account for informative terminal events, such as premature treatment discontinuation or death, which prevent the occurrence of subsequent events. To address these limitations, we derive and validate estimation and inference procedures for the area under the mean cumulative function (AUMCF), an extension of the restricted mean survival time to the multiple event-time setting. The AUMCF is clinically interpretable, properly accounts for terminal competing risks, and can be estimated nonparametrically. To enable covariate adjustment, we also develop an augmentation estimator that provides efficiency at least equaling, and often exceeding, the unadjusted estimator. The utility and interpretability of the AUMCF are illustrated with extensive simulation studies and through an analysis of multiple heart-failure-related endpoints using data from the Beta-Blocker Evaluation of Survival Trial (BEST) clinical trial. Our open-source \texttt{R} package \texttt{MCC} makes conducting AUMCF analyses straightforward and accessible. \\
\end{abstract}

%  Please place your key words in alphabetical order, separated
%  by semicolons, with the first letter of the first word capitalized,
%  and a period at the end of the list.
%

\maketitle
%  As usual, the \maketitle command creates the title and author/affiliations
%  display 

%  If you are using the referee option, a new page, numbered page 1, will
%  start after the summary and keywords. The page numbers thus count the
%  number of pages of your manuscript in the preferred submission style.
%  Remember, ``Normally, regular papers exceeding 25 pages and Reader Reaction 
%  papers exceeding 12 pages in (the preferred style) will be returned to 
%  the authors without review. The page limit includes acknowledgements, 
%  references, and appendices, but not tables and figures. The page count does 
%  not include the title page and abstract. A maximum of six (6) tables or 
%  figures combined is often required.''

%  You may now place the substance of your manuscript here. Please use
%  the \section, \subsection, etc commands as described in the user guide.
%  Please use \label and \ref commands to cross-reference sections, equations,
%  tables, figures, etc.
%
%  Please DO NOT attempt to reformat the style of equation numbering!
%  For that matter, please do not attempt to redefine anything!

\section{Introduction}
\label{s:intro}
Multiple event-time data arise in comparative studies when patients experience more than one non-fatal event over time \citep{glynn2001counting, rogers2012eplerenone, rogers2014analysing, claggett2018quantifying, mogensen2018effect, solomon2018baseline, pfeffer2022impact}. These events may be repeated events of the same type, as in recurrent events data, or encompass several distinct types of events, which we refer to as multiple event data. In chronic disease trials, improvements in care have led to longer follow-up and an increased likelihood of patients experiencing multiple events. Our motivating example is the Beta-Blocker Evaluation of Survival Trial (BEST) \citep{best1995, best2001}. BEST was a randomized, double-blind, placebo-controlled trial evaluating the effect of bucindolol hydrochloride, a non-selective beta-blocker and mild vasodilator, on patients with advanced heart failure and reduced ejection fraction. Although the primary endpoint was overall survival, the trial captured several secondary outcomes to provide a more comprehensive understanding of disease burden. These included hospitalization, myocardial infarction, and heart transplant.

To fully leverage information from repeated events, a variety of statistical methods have been developed for analyzing multiple and recurrent event data. Early work modeled recurrent events as non-homogeneous Poisson processes and relied on proportional intensity assumptions to incorporate the effects of baseline covariate (e.g., \cite{prentice1981regression, lawless1987regression}). A second line of work modeled the marginal rate function (e.g., \cite{wei1989regression, pepe1993some, lawless1995some}), but these approaches generally require that events are not interrupted by terminal events such as death. For example, \cite{lin2000semiparametric} proposed a regression-based approach, termed ``LWYY'', which also assumes constant proportional covariate effects on the baseline mean frequency function.

To account for the presence of intercurrent terminal events, marginal (e.g., \cite{cook1997marginal, li1997use, zhao2013marginal}) and frailty models (e.g., \cite{wang2001analyzing, huang2004joint, han2020variable}) have been proposed to accommodate the dependence between recurrent and terminal events. However, most methods impose restrictive assumptions on the event process and/or the dependence of the recurrent and terminal event processes that may not hold in practice. In response, several nonparametric methods have been introduced. \cite{sparapani2020non} introduced Bayesian Additive Regression Tree (BART) methods for analyzing recurrent events data with baseline covariates. In a separate line of work, \cite{ghosh2000nonparametric} proposed to estimate the mean cumulative function (MCF) of a multiple or recurrent event process. The MCF is the marginal mean of the cumulative number of events over time. Terminal events are accomodated by assuming that the cumulative number of events remains constant after the terminal event. A formal two group comparison based on the estimated MCFs that mimics the familiar log-rank test was developed in \cite{ghosh2002marginal}. However, the testing approach was not accompanied by a clinically interpretable summary measure of treatment efficacy.

Recently, the area under the mean cumulative function (AUMCF) has been used as such a summary \citep{claggett2022nejm}. The AUMCF represents the expected total time lost to undesirable events. Simply put, the higher the MCF, the larger the AUMCF and the less effective the treatment. The ratio or difference in AUMCFs across study arms provides a natural measure of how treatment influences the cumulative disease burden experienced by patients over the study period. A review of 5 cardiovascular clinical trials \citep{gregson2023} found that the AUMCF required fewer assumptions, but reached similar conclusions to several common comparators, including the Cox proportional hazards model, negative binomial regression, the LWYY method, frailty models, and the win ratio. The win ratio and related statistics, such as the win loss \citep{luo2017weighted}, win odds \citep{brunner2021win}, net benefit \citep{verbeeck2020evaluation}, and last-event-assisted win ratio \citep{mao2022recurrent}, have become particularly popular for recurrent or hierarchical multiple event data. While appealing for their ability to accommodate events of differing severity or priority, win statistics can be difficult to interpret and are sensitive to the chosen ordering of events \citep{pocock2012win, pocock2024win}. In contrast, the AUMCF is a conceptually straightforward summary of total event burden over time and avoids the need to prioritize events or model complex dependence structures.

However, the theoretical foundations of the AUMCF have yet to be fully developed, which are essential for its complete validation. To fill this gap, we formally derive and rigorously validate a nonparametric method for estimating the AUMCF. In Section 2, we introduce the proposed estimation and inference procedures, including an approach for baseline covariate adjustment. In Section 3, we apply our method to an analysis of multiple event-time data from BEST and evaluate the operating characteristics of our proposal with extensive simulation studies. We conclude with final remarks in Section 4.

% -------------------------------------------------

\section{Method}\label{methods}
\subsection{Problem Setting and Notation}
Our primary focus is on a comparative study in which multiple events as well as a terminal event can occur over an observation period of length $\tau$. We let $D$ denote the time of the terminal event, $C$ the censoring time, and $\bm{W}$ a $p$-dimensional vector of baseline covariates. The latent underlying counting process for the events is defined as $N_0(t) = \sum_{l=1}^{\infty} \mathbb{I}(T_l \le t)$, where $T_l$ is the time of the $l^\text{th}$ event and $\Ibb(\cdot)$ is the indicator function. In the presence of a terminal event, the counting process cannot jump after $D$. Letting $a \wedge b = \min(a,b)$, we denote the truncated-by-death process as $N^*(t) = N_0(t \wedge D) =\int_{0}^t  \mathbb{I}(D > u) dN_0(u)$. We assume that $C$ is independent of $(D, {N}^*(\cdot), \bm{W}^T)^T$ within each treatment arm, but do not impose any assumptions on the dependency of the event times or $D$ and ${N}^*(t)$. 

Due to censoring, a patient is only observed until time $X = D \wedge C$. We let $\delta = \mathbb{I}(D \le C)$ and $N(t) = {N}^*(t \wedge C)$ be the observed counting process. The observable data therefore consist of independent observations 
$\{ (N_{ij}(\cdot), X_{ij}, \delta_{ij}, \bm{W}_{ij}^T)^T \mid i = 1, \dots, n_j\},$
where $i$ indexes the subject and $j$ indexes the treatment arm. We define the total sample size as $n = n_1 + n_2$ and assume that $\lim_{n \to \infty}   n_j / n = \rho_j \in (0,1)$ for $j = 1,2$. Additionally, we define the at-risk process as $Y_{ij}(t) = \mathbb{I}(X_{ij} \ge t)$, which indicates whether subject $i$ in arm $j$ is under observation and at risk for events at time $t \in (0, \tau]$. For simplicity, we suppress individual and arm indices in later sections when doing so does not lead to ambiguity.

% -------------------------------------------------

\subsection{The Area Under the Mean Cumulative Function (AUMCF)} 

\subsubsection{One sample setting}
To motivate our thinking, we first consider the setting of a single treatment arm without a terminal event or baseline covariates. In this setting, the mean cumulative function (MCF) is $m(t) := \mathbb{E} \left\{ {N}_0(t) \right\}$. The MCF is an intuitive summary of the counting process as it is the expected number of events by time $t \in (0, \tau]$. We propose to summarize the expected disease burden of patients with the area under the MCF (AUMCF), defined as 
$\theta := \int_0^\tau m(t)dt$. To understand this estimand, note that the AUMCF can be expressed as $\theta =\mathbb{E}\left\{\sum_{l=1}^{\infty}  \left(\tau - T_l \right)_+\right\},$
where $(x)_+=x \Ibb(x>0)$. The AUMCF is therefore the total event-free time lost from all events and may be interpreted clinically as the disease burden experienced by participants during the study. 

In the presence of a terminal event, the MCF is $ m(t) := \mathbb{E} \left\{ {N}^*(t) \right\} = \int_0^t S_D(u) dR(u)$ where $S_D(t) = \Pbb(D \ge t)$ is the survival function of $D$ and $dR(t) = \mathbb{E}\{d N^*(t) \mid D \ge t\}$ \citep{ghosh2000nonparametric}. The AUMCF is then defined as
\begin{align}\label{theta-terminal}
     \theta  := \int_0^\tau \left\{ \int_0^t S_D(u) dR(u) \right\}  dt = \int_0^\tau (\tau - u) S_D(u) \, dR(u). 
\end{align}
While the MCF and AUMCF maintain the same straightforward interpretation as the setting without a terminal event, the key difference is that additional recurrences of the event(s) of interest cannot take place after the terminal event. Figure \ref{fig:combined_overview} provides intuition for the calculation of the AUMCF in the presence of a terminal event. 

It follows from \cite{ghosh2000nonparametric} that $m(t)$ can be estimated with
$\widehat{m}(t)=\int_0^t \widehat{S}_D(u)\frac{d \bar{N}(u)}{\bar{Y}(u)}$
where $\bar{N}(t) = \sum_{i=1}^n {N}_i(t)$ and $\bar{Y}(t)=\sum_{i=1}^n Y_{i}(t) $ are the sample-level counting processes and $\widehat{S}_D(t)$ is the Kaplan-Meier estimator of ${S}_D(t)$. Note that $R(t)$ is estimated by the familiar Nelson-Aalen estimator and therefore $\widehat m(t)$ is a plug-in estimator of $m(t)$. We correspondingly propose to estimate $\theta$ in (\ref{theta-terminal}) as
\begin{align}\label{eq: one_samp_est}
    \widehat{\theta}=\int_0^\tau (\tau-u)\widehat{S}_D(u)\frac{d \bar{N}(u)}{\bar{Y}(u)}.
\end{align}

\remark{Suppose that only the terminal event is of interest, such that $N^{*}(t)$ is the counting process for death and only death. Then $dR(t) = \Ebb\{dN^{*}(t)| D \geq t\} = \Pbb\{D \in [t, t+dt) | D \geq t\}$,
and the MCF reduces to the cumulative incidence of death,
\begin{align*}
    m(t) = \int_{0}^{t}S_{D}(u)dR(u) = \int_{0}^{t}S_{D}(u)\frac{dF_{D}(u)}{S_{D}(u)} = F_{D}(t).
\end{align*}
In this setting, the AUMCF is $\theta = \int_{0}^{\tau}F_{D}(t)dt$, 
which is interpretable as the event-free time lost due to death. Substituting $1-S_{D}(t)$ for $F_{D}(t)$ yields $\theta = \tau - \int_{0}^{\tau}S(t)dt$, where $\int_{0}^{\tau}S_{D}(t)dt$ is the familiar restricted mean survival time (RMST). The RMST has the simplified interpretation as the total amount of time a subject survives, on average, up to time $\tau$ \citep{uno2014rmst, mccaw2019rmst}.

\remark{ If a treatment reduces the rate of the terminal event, but otherwise has no effect on the recurrence rate, it is possible for the MCF,  and hence the AUMCF, to increase by extending the time during which patients are at-risk to experience recurrences. In this scenario, direct comparison of the MCFs or AUMCFs may be misleading. We investigate the empirical impact of such survival bias in Web Appendix B.2. }

\begin{figure}[htbp]
    \centering
    \begin{subfigure}{\textwidth}
        \centering
        \includegraphics[scale = 0.45]{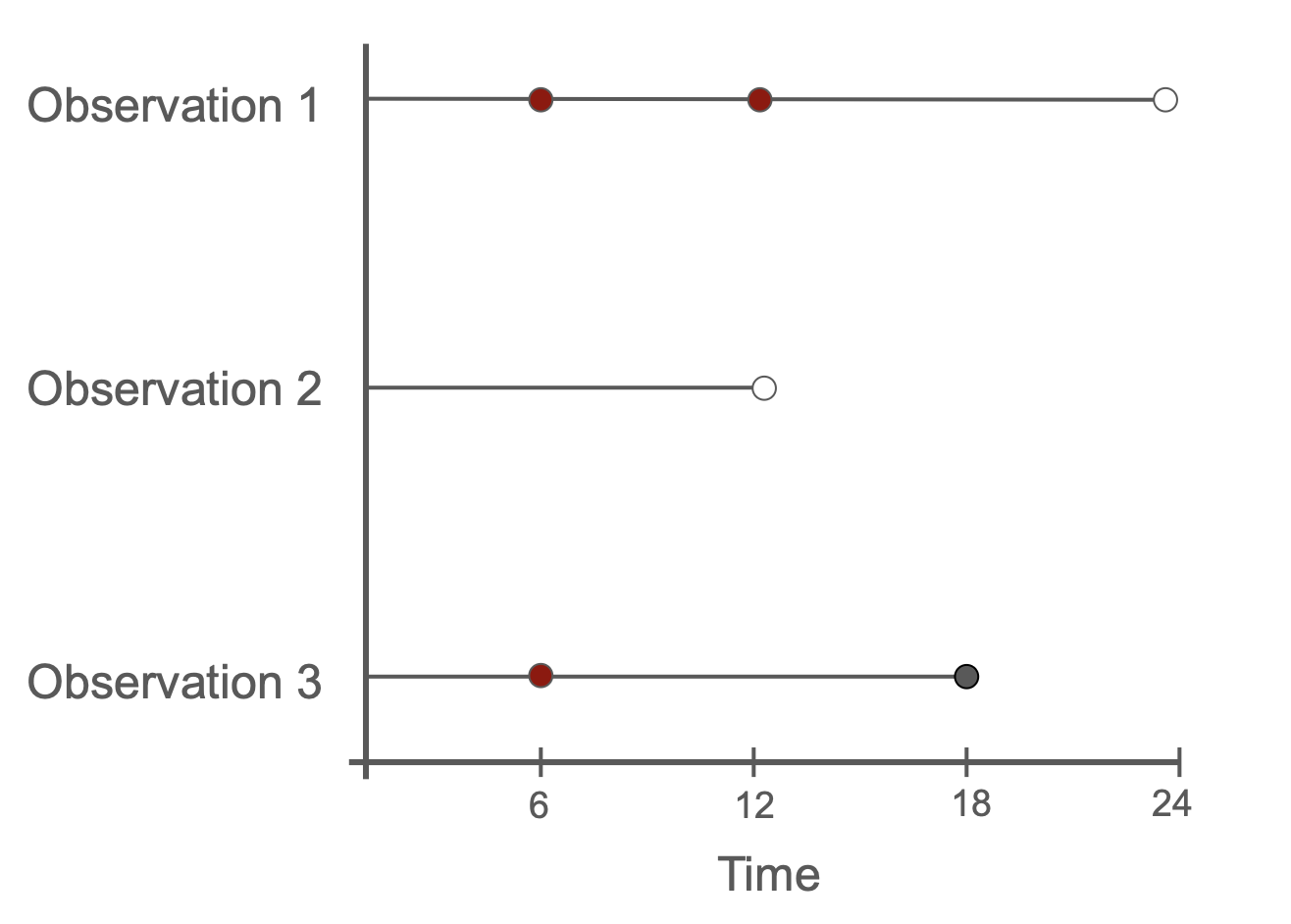}
    \end{subfigure}
    \hfill
    \begin{subfigure}{\textwidth}
        \centering
        \includegraphics[width=\textwidth]{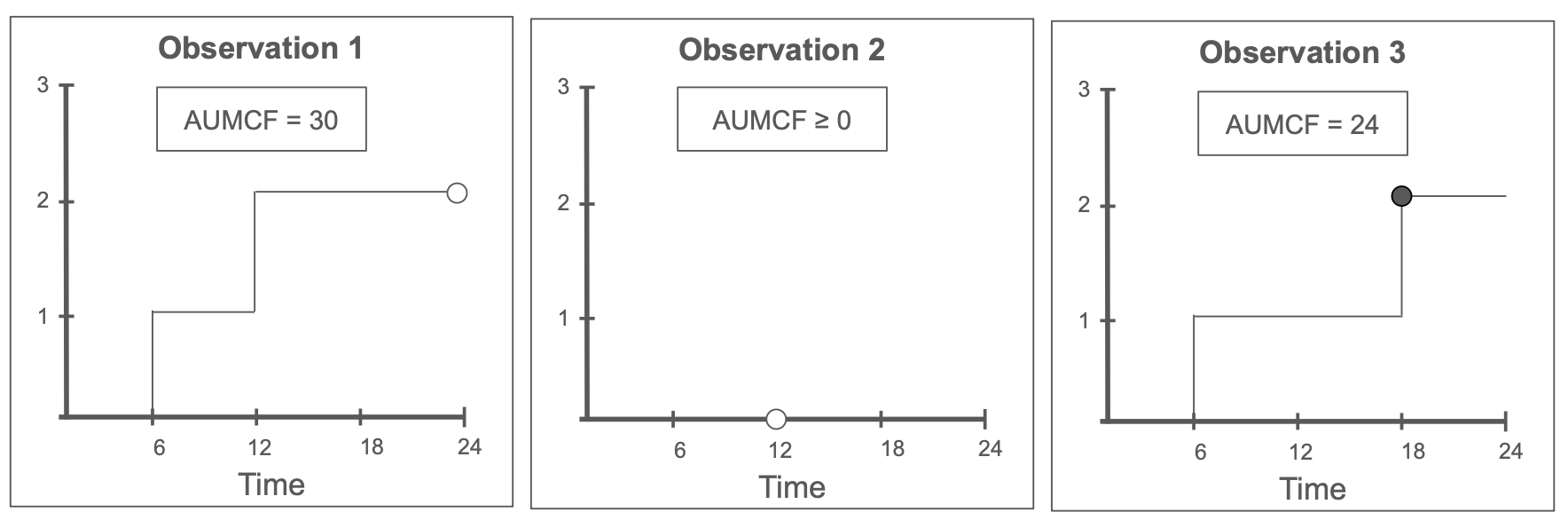}
    \end{subfigure}
    \caption{ {\bf{Overview of the area under the mean cumulative function (AUMCF).}} The AUMCF quantifies the total event-free time lost from all events and is illustrated in an example of a study with a 24 month follow-up period. The top panel shows three hypothetical observations while the bottom three panels illustrates the corresponding observation-level MCFs and AUMCFs. In observation 1, two non-terminal events (red circles) are experienced at months 6 and 12 and the process is censored at the end of follow-up (white circle). Therefore, 18 and 12 event-free months were lost and the AUMCF is equal to (24-6) + (24-12) = 30 event months. In observation 2, no events are experienced during 12 months before censoring occurs, which could represent late enrollment into the study. All that is known is that the AUMCF is positive since no event-free time is lost. In observation 3, one non-terminal event occurs at 6 months and the terminal event (black circle) occurs at 18 months of follow-up. The AUMCF is equal to (24-6) + (24-18) = 24 event months.}
    \label{fig:combined_overview}
\end{figure}

% -------------------------------------------------

\subsubsection{Two sample setting}
In randomized controlled trials, interest typically lies in comparing event rates between two treatment arms. In this case, the MCF and the AUMCF are respectively defined for each treatment group as 
\begin{align*}
    m_{j}(t) := \mathbb{E}\{{N}^*_j(t)\} \quad \mbox{and} \quad \theta_{j} :=\int_0^\tau m_{j}(t)dt \quad \mbox{for} \quad j = 1, 2
\end{align*}
where ${N}^*_j(t)$ is the underlying counting process in the $j$th treatment arm. To compare the event rates of two arms, we focus on estimation of the difference in AUMCFs, defined as $\Delta =\theta_{1}-\theta_{2}$. Using the estimator of the AUMCF proposed in (\ref{eq: one_samp_est}), we estimate $\Delta$ as
\begin{align}
    \widehat{\Delta} =\widehat{\theta}_{1}-\widehat{\theta}_{2} = \int_0^\tau (\tau-u)\left\{\frac{\widehat{S}^D_{1}(u)d \bar{N}_1(u)}{\bar{Y}_1(u)}-\frac{\widehat{S}^D_{2}(u)d \bar{N}_2(u)}{\bar{Y}_2(u)} \right\}
    \label{eqn:estimated-delta}
\end{align}
where
$\bar{N}_j(u)=\sum_{i=1}^{n_j} N_{ij}(u),$ $\widehat{S}_{j}^{D}(u)$ is the Kaplan-Meier estimator of the survival function of $D_j$, and $\bar{Y}_j(u)=\sum_{i=1}^{n_j} \Ibb(X_{ij}\ge u)$ for $j = 1,2$. Inference on $\widehat{\Delta}$ can be performed based on its asymptotic normality, which is presented in Theorem \ref{thm: delta-IF} and formally derived in Web Appendices A.1 and A.2.

\begin{theorem}\label{thm: delta-IF}
Let $\Lambda_j^D(t)$ be the cumulative hazard function of the terminal event in the $j$th treatment arm, $N_{ij}^D(t) = \mathbb{I}( X_{ij} \le t, \delta_{ij} = 1)$, $M_{ij}^D(t) = N_{ij}^D(t) - \int_0^t Y_{ij}(u) d\Lambda_j^D(u)$, $M_{ij}(t) =  N_{ij}(t) - \int_{0}^t Y_{ij}(u) dR_j(u)$, and $\pi_j =  \mathbb{P}( X_j \ge t)$. Then, $\left(\dfrac{n_1 n_2}{n}\right)^{1/2}( \hat{\Delta} - \Delta)$ converges weakly to a mean-zero normal distribution with variance $\Sigma_\Delta = \rho_2 \Sigma_1 + \rho_1\Sigma_2$ 
 where $\Sigma_j = \mathbb{E}( \Psi_{ij}^2)$ and  $\Psi_{ij} =  \int_0^\tau (\tau-u)S_{_j}^{D}(u)\frac{dM_{ij}(u)}{\pi_j(u)} - \int_0^\tau  \left\{ \int_u^\tau  (\tau-v)  dm_j(v) \right\}  \frac{dM_{ij}^{D}(u)}{\pi_j(u)}.$

\end{theorem}
Large-sample confidence intervals and hypothesis tests for $\Delta$ can be obtained with a consistent estimator of $\Sigma_\Delta$ constructed as  $\widehat{\Sigma}_\Delta =  \frac{n_2}{n}\widehat{\Sigma}_{1} + \frac{n_1}{n}\widehat{\Sigma}_2$ where $\widehat{\Sigma}_j = n_j^{-1} \sum_{i =1 }^{n_j} \widehat{\Psi}_{ij}^2$ and $\widehat{\Psi}_{ij}$ is obtained by replacing the unknown quantities in $\Psi_{ij}$ with their respective estimators. A complete expression for $\widehat{\Sigma}_\Delta$ is provided in Web Appendix A.3.

\remark{A test statistic mirroring the log-rank test statistic for recurrent events analysis was proposed in \cite{ghosh2000nonparametric} and takes the form
\begin{align*}
    Q=\int_0^\tau \frac{(n_1n_2)^{-1}\bar{Y_1}(u)\bar{Y}_2(u)}{n^{-1}\{\bar{Y}_1(u)+\bar{Y}_2(u)\}}\left[\frac{\widehat{S}_{D_1}(u)d \bar{N}_1(u)}{\bar{Y}_1(u)}-\frac{\widehat{S}_{D_2}(u)d \bar{N}_2(u)}{\bar{Y}_2(u)} \right].
\end{align*}
Under a general alternative, $Q$ converges in probability to a non-zero limit that depends on the censoring distributions in each treatment arm. In contrast, the estimator of $\widehat{\Delta}$ in (\ref{eqn:estimated-delta}) is always consistent for $\Delta$ and the Wald-test statistic based on $\widehat{\Delta}$ is consistent with the corresponding Wald-type confidence interval. That is, a $p$-value for the test $H_0: \Delta = \Delta_0$ less than the significance level, $\alpha$, implies the $100(1-\alpha)$\% confidence interval for $\Delta$ will exclude $\Delta_0$ and vice versa. The test based on $Q$ does not generally possess this intuitive property.}

% -------------------------------------------------
\subsubsection{Covariate adjustment}
\label{sec:covar-adjust}
Our discussion so far has not incorporated baseline covariates. A substantial body of literature underscores the value of covariate adjustment in randomized controlled trials to enhance precision (e.g., \cite{senn1989covariate, tsiatis2008covariate}). Building on regression-based adjustment, \cite{zhang2008improving} proposed a general and flexible framework grounded in semiparametric theory to improve the efficiency of unadjusted treatment effect estimators. This approach has been used across a range of settings (e.g., \cite{tian2012covariate, parast2017evaluating, gronsbell2024another}), and we adopt it here to construct a covariate-adjusted estimator of the AUMCF in the two-sample setting. 

The underlying idea of our approach is to augment the unadjusted estimator with a carefully constructed term that captures potential covariate imbalance between treatment arms. More formally, we propose the augmented estimator $\widehat{\Delta} - {\bm \beta}^T \left( \overline{\bm W}_1 - \overline{\bm W}_2\right)$ where $\overline{\bm W}_j = n_j^{-1} \sum_{i = 1}^{n_j} {\bm W}_{ij}$ and ${\bm \beta}$ is a weight vector.  Due to randomization, the augmentation term, $\left( \overline{\bm W}_1 - \overline{\bm W}_2\right)$, converges to zero in probability so that the estimator retains consistency for $\Delta$. To minimize the asymptotic variance of the augmented estimator, we propose to take the weight as $\bm{\beta}_0 = \Sigma_{\bm{W}}^{-1} {\bm \gamma} _0$ where $\Sigma_{\bm{W}} =  \rho_2 \text{Var}(\bm{W}_{i1}) +  \rho_1 \text{Var}(\bm{W}_{i2})$ and ${\bm \gamma} _0 = \rho_2  \text{Cov}(\bm{W}_{i1}, \Psi_{i1}) +  \rho_1  \text{Cov}(\bm{W}_{i2}, \Psi_{i2})$. We estimate $\bm{\beta}_0$ with $\widehat{\bm \beta}=\widehat{\Sigma}_{\bm{W}}^{-1}\widehat{\bm \gamma}$ where 
\begin{align*}
    \widehat{\bm \gamma}= \left(\frac{n_2}{n}\right) \frac{1}{n_1} \sum_{i=1}^{n_1}  \left(\bm{W}_{i1}-\overline{\bm W}_1\right) \widehat{\Psi}_{i1} + \left(\frac{n_1}{n}\right) \frac{1}{n_2} \sum_{i=1}^{n_2}  \left(\bm{W}_{i2}-\overline{\bm W}_2\right) \widehat{\Psi}_{i2}
\end{align*}
and
\begin{align*}
    \widehat{\Sigma}_{\bm{W}}= \left(\frac{n_2}{n}\right) \frac{1}{n_1 }\sum_{i=1}^{n_1} \left(\bm{W}_{i1}-\overline{\bm W}_1\right) \left(\bm{W}_{i1}-\overline{\bm W}_1\right)^T + \left(\frac{n_1}{n}\right) \frac{1}{n_2 }\sum_{i=1}^{n_2} \left(\bm{W}_{i2}-\overline{\bm W}_2\right) \left(\bm{W}_{i2}-\overline{\bm W}_2\right)^T.
\end{align*}
The final augmented estimator of $\Delta$ is then taken as  $\widehat{\Delta}^{\aug}= \widehat{\Delta} - \widehat{\bm \beta}^T \left( \overline{\bm W}_1 - \overline{\bm W}_2\right)$. We detail the asymptotic properties of $\widehat{\Delta}^{\aug}$ in Web Appendix A.3 and summarize our main result in the following theorem.    
\begin{theorem}\label{thm: cov-adj}
$\hat{\Delta}^{\aug}$ is consistent for $\Delta$ and $\left(\dfrac{n_1 n_2}{n}\right)^{1/2}( \hat{\Delta}^{\aug} - \Delta)$ converges weakly to a mean-zero normal distribution with variance $\Sigma_\Delta - {\bm \gamma}_0^T  {\Sigma}_{\bm{W}}^{-1}  {\bm \gamma}_0$.  
\end{theorem}
By Theorems \ref{thm: delta-IF} and \ref{thm: cov-adj}, the variance of $\widehat{\Delta}^{\aug}$ is no greater than that of $\widehat{\Delta}$ and  the improvement in precision depends on the degree of covariate imbalance.

% -------------------------------------------------
\section{Empirical studies}
To illustrate the practical utility of the AUMCF, we first return to the Beta-Blocker Evaluation of Survival Trial (BEST) and then evaluate the operating characteristics of the AUMCF through extensive simulation studies. For both sets of analyses, we compared the performance of the proposed estimator for the difference in AUMCFs with six commonly used methods: the Cox proportional hazards model, negative binomial regression, the shared frailty model (Frailty), the LWYY method (LWYY), the standard win ratio (SWR), and the last-event-assisted win ratio (LWR). Inference for the LWYY method is based on the robust sandwich variance estimate. These methods were selected based on a recent review of approaches for multiple event time data \citep{gregson2023}.}

\subsection{Real data example}
In BEST, a total of 2,708 eligible patients were randomized 1:1 to bucindolol or placebo. The primary endpoint was overall survival (OS) and the secondary endpoints include the times to hospitalization (from heart failure or any cause), myocardial infarction, and heart transplantation. Based on an observed hazard ratio (HR) of 0.90 (95\% CI, 0.78 to 1.02, $P=0.11$) for OS in the primary analysis, the study concluded there was no significant evidence of an OS benefit in the overall study population. We reanalyze data from BEST not to reach different conclusions, but to illustrate how to apply the AUMCF methodology.

Our data set includes 1,354 patients randomized to bucindolol and 1,353 randomized to placebo. For the primary endpoint of OS, there were 411 events in the bucindolol arm, and 448 events in the placebo arm. Kaplan-Meier (KM) estimates of the OS probability across 48 months of follow-up are presented in \textbf{Web Figure 3}. The 48-month RMST, which is the area under the Kaplan-Meier curve across the follow-up period, was 36.15 months for bucindolol and 35.10 months for placebo. This means that a heart failure patient randomized to bucindolol and followed for 48 months would expect to survive 36.15 months. The difference in 48-month RMSTs was 1.05 months (95\% CI, -0.29 to 2.4 months, $P = 0.13$) in favor of bucindolol. 

OS may not capture disease burden as indicated, for example, by a hospitalization. We defined a composite disease-burden endpoint as the time to first hospitalization from any cause, myocardial infarction, heart transplantation, or death. For this endpoint, there were 931 events in the bucindolol arm versus 971 in the placebo arm. Kaplan-Meier curves for event-free survival across the 48 months of follow-up are presented in \textbf{Web Figure 4}. The 48-month RMSTs were 19.1 months for bucindolol and 18.0 months for placebo. Here, the 19.1 months is the expected event-free survival time of a patient randomized to bucindolol and scheduled for 48 months of follow-up. The difference in 48-month RMSTs was 1.16 months (95\% CI, -0.21 to 2.5 months, $P = 0.098$) in favor of bucindolol while the HR with respect to the composite endpoint was 0.94 (95\% CI, 0.85 to 1.02, $P=0.15$), and the SWR was 0.95 (95\% CI, 0.70 to 1.30, $P=0.76$).

By disregarding events following the first event, time-to-first event analysis makes inefficient use of the data, ignores the co-occurrence of events within patients, and may understate a patient's full disease burden. This motivates the consideration of multiple event-time endpoints, which incorporate the timing and occurrence of all heart failure–related events and thereby provide a more comprehensive picture of a patient's disease trajectory. In BEST, 833 patients in the bucindolol arm and 870 patients in the placebo arm experienced more than 1 event. \textbf{Table \ref{tab:best-analysis-results}} presents results from multiple event-time analysis of the composite disease-burden endpoint via several methods, taking into account all occurrences. It should be noted that these methods differ in their underlying assumptions and target estimands, hence the point estimates and statistical inferences need not agree. Nonetheless, all methods are directionally consistent in suggesting a benefit from bucindolol, and all methods except the last-event win-ratio are nominally significant at the $\alpha = 0.05$ level.

\begin{figure}[ht]
    \centering
    \includegraphics[width=\textwidth]{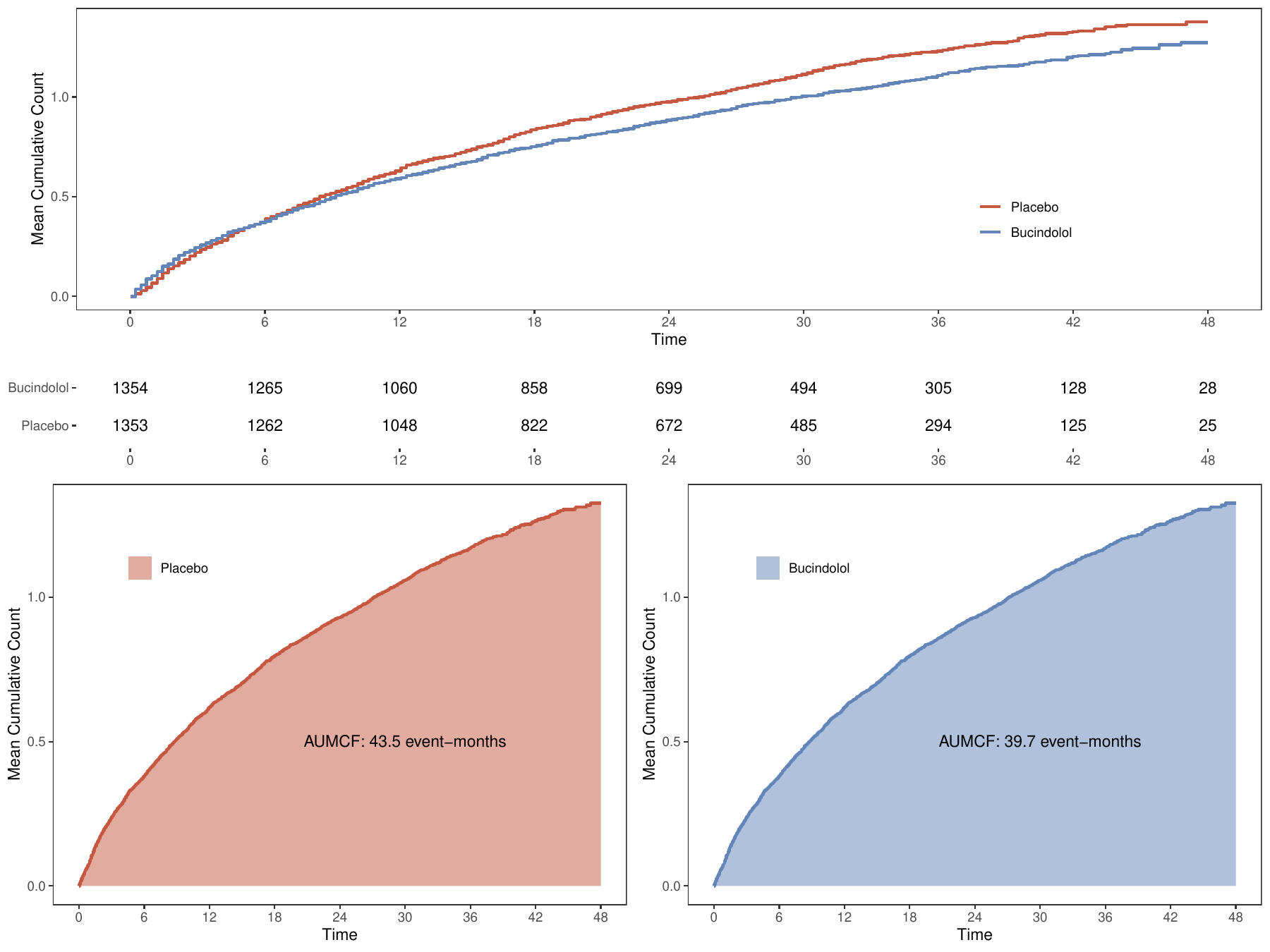}
    \caption{\textbf{Mean cumulative functions (MCF) and area under the MCF (AUMCF) for patients from the Beta-Blocker Evaluation of Survival (BEST) Trial.} The composite disease-burden endpoint consists of the times to hospitalization from any cause, myocardial infarction, heart transplantation, and death, considering all occurrences a patient experienced.}
    \label{fig:best-mcc}
\end{figure}

The AUMCF analysis is illustrated in \textbf{Figure \ref{fig:best-mcc}}. Across 48-months of follow-up, the AUMCFs were 39.7 months for bucindolol and 43.5 months for placebo. Note the difference in interpretation compared with the RMST: the AUMCF is the expected event-free time \textit{lost} across the follow-up period. Thus, a lower value is better. The difference in 48-month AUMCFs was 3.73 months (95\% CI, 1.32 to 6.15 months, $P = 0.0025$) in favor of bucindolol. That is, across the 48 months of follow-up, patients randomized to bucindolol survived event-free (i.e.\@, free of hospitalization, myocardial infarction, or heart transplanation) for an additional 3.73 months on average. 

To illustrate the use of our augmentation estimator, we also conducted an analysis adjusting for baseline age, sex, left ventricular ejection fraction, and an indicator for NYHA class IV heart failure. The augmented estimate of the treatment difference was 3.96 months (95\% CI, 1.62 to 6.31, $P = 9.4 \times 10^{-4}$). In addition to increasing the magnitude of the estimated treatment difference, covariate adjustment reduced the estimated standard error from 1.240 to 1.202, yielding a 6.3\% improvement in relative efficiency. We caution that the modest gain in efficiency observed in the BEST trial should not be taken to imply that the augmentation estimator will offer limited benefit in general. In a large, well-powered, randomized clinical trial like BEST, minimal covariate imbalance is expected, leaving little room for further improvements in efficiency. Results from additional sensitivity analyses are presented in the Web Appendix C. 

\begin{table}[ht]
    \centering
    \begin{tabular}{c c c c}
    \hline
    Method & Contrast Estimate & 95\% CI & P-value \\ \hline
    AUMCF & 3.73 & $(1.32, 6.15)$ & 0.0025 \\
    Frailty & 0.89 & $(0.83, 0.96)$ & 0.0034 \\
    LWYY & 0.89 & $(0.83, 0.96)$ & 0.0034 \\
    NB & 0.89 & $(0.83, 0.97)$ & 0.0059 \\
    LWR & 1.09 & $(0.99, 1.20)$ & 0.069 \\ \hline
    \end{tabular}
    \bigskip
    \caption{\textbf{Multiple event-time analyses of data from the Beta-Blocker Evaluation of Survival (BEST) Trial.} The composite disease-burden endpoint consists of the times to hospitalization from any cause, myocardial infarction, heart transplantation, and death, considering all occurrence a patient experienced. Each statistic contrasts the bucindolol and placebo arms. The difference in the AUMCF across treatment arms is compared with the shared frailty model (Frailty), the LWYY method (LWYY), negative binomial regression (NB), and the last-event-assisted win ratio (LWR).}
    \label{tab:best-analysis-results}
\end{table}

\subsection{Simulation studies}
\subsubsection{Data generation}
We considered three scenarios by generating data under (i) an independent competing risks model, (ii) a frailty model, or (iii) a time-vary treatment effect model.  Data were simulated from two independent groups with a sample size per arm of $n_j \in \{100, 200, 400\}$ for $j = 1, 2$. In scenario (i), recurrent events in arm $j$ were simulated from a homogeneous Poisson process with event rate $\lambda_{E_j}$ and terminal events were simulated independently from an exponential distribution with rate $\lambda_{D_j}$. In scenario (ii), recurrent events in arm $j$ were simulated from a homogeneous Poisson process with rate $\lambda_{E_j} \times \xi_j$ and terminal events were simulated from an exponential distribution with rate $\lambda_{D_j} \times \xi_j$ where $\xi_j$ is a frailty component following a Gamma distribution with mean 1 and variance 3. In scenario (iii), recurrent events were simulated from a non-homogeneous Poisson process with rate $\lambda_{E_j}$ up to time $c$ and rate $\upsilon_j \lambda_{E_j}$ after time $c$ for $c \in (0, \tau)$ while terminal events were simulated from an exponential distribution with rate $\lambda_{D_j}$. In all scenarios, event-times were subject to independent censoring and the censoring time was generated from an exponential distribution with rate $\lambda_{C} = 0.2$.

We evaluated settings under the null and alternative hypotheses for each scenario. For scenarios (i) and (ii), $(\lambda_{E_1}, \lambda_{E_2}) = (1, 1)$ under the null and $(\lambda_{E_1}, \lambda_{E_2}) = (1.4, 1)$ under the alternative. For scenario (iii), $(\lambda_{E_1}, \lambda_{E_2}) = (1, 1)$ and $(\upsilon_1, \upsilon_2) = (0.5, 0.5)$ under the null and $(\lambda_{E_1}, \lambda_{E_2}) = (1, 1)$ and $(\upsilon_1, \upsilon_2) = (1, 0.5)$ under the alternative. We let $(\lambda_{D_1}, \lambda_{D_2}) = (0.2, 0.2)$ in all settings. To evaluate the performance of the covariate adjustment strategy, we also considered a special setting of scenarios (i) and (ii) under the null using a baseline covariate, $W$, simulated for all subjects from a standard normal distribution. Two settings were considered, one in which the covariate is uninformative, having no effect on either the event or death rates, and a second in which it is informative. For the latter, each subject's death rate was scaled by $\exp\{ W \times\ln(0.5) \}$ and each subject's event rate was scaled by $\exp\{ W \times\ln(2) \}$. We considered observation lengths of $\tau \in \{1, 4\}$ for
scenarios (i) and (ii). For scenario (iii), we considered $\tau \in \{4, 6\}$ and let $c = 1$. All results in the main text are summarized across 10,000 simulation replicates.

\subsubsection{Evaluation metrics}
For all scenarios, we evaluated the performance of point and standard error estimation with the bias and the asymptotic standard error (ASE), respectively. We also assessed type I error in the null settings and power in the alternative settings. For all methods except the SWR and the LWR, we evaluated coverage probability (CP) of 95\% confidence intervals (CIs) in the alternative settings. The true values of each parameter were calculated as the average of 2,000 estimates from a dataset generated under no censoring and with 10,000 observations in each arm. As the estimands for the SWR and LWR generally depend on the censoring distribution, we do not present CP for these methods \citep{mao2019alternative, mao2024defining}. We focus on the setting of $n_1 = n_2 = 200$ and provide results for the additional sample sizes in Web Appendix B.1.

\subsubsection{Results}
\textbf{Table \ref{tab:c1ac2_n200}} summarizes the operating characteristics of our proposed method and the six comparators under the independent competing risks and the frailty models, corresponding to scenarios (i) and (ii), respectively. Under the null in the independent competing risks scenario, all methods exhibit negligible bias, provide accurate standard error estimates, and maintain appropriate type I error control. Under the alternative, the Cox proportional hazards model has substantially lower power than the AUMCF, as it only considers the time to the first event. In contrast, the comparable parametric methods that accommodate multiple events, namely the frailty model, the LWYY method, and negative binomial regression, achieve higher power than the AUMCF, particularly with the shorter observation window, as their underlying assumptions are satisfied in this setting. The last-event assisted win ratio and the standard win ratio have substantially lower power than all of the other methods. The power is the lowest in the setting with the longer observation window as win ratio methods generally prioritize early events. While the parametric methods exhibit inflated standard error estimation, CP is generally close to the nominal level for the 95\% CIs for all of the methods.

For the frailty model scenario, the results are similar under the null, with the one exception being that the negative binomial regression exhibits slightly inflated type I error. Under the alternative, findings also mirror those from the independent competing risks scenario, with the exception that the AUMCF achieves comparable power to the LWYY method and negative binomial regression and higher power than the frailty model. The results for $n_1 = n_2 = 100$ and 400 are presented in \textbf{Web Tables 1} and \textbf{3} and generally follow the same patterns. Notably, the LWYY and negative binomial models exhibit inflated type I error under the frailty model scenario for the larger sample size.

\textbf{Table \ref{tab:c3_n200}} summarizes results for the time-varying treatment effect model of scenario (iii). Under the null, all methods maintain accurate point and standard error estimation and hence proper type I error control. Under the alternative, only the AUMCF retains these properties while also achieving high power. In contrast, the comparable parametric methods exhibit substantial upward bias (relative to the ESE), leading to deceptively high power and poor coverage of the 95\% CIs, particularly when the observation window is longer. Although the LWR and SWR methods yield accurate point and standard error estimates, their power is even lower than that of the Cox proportional hazards model. Results for $n_1 = n_2 = 100$ and 400, presented in \textbf{Web Tables 2} and \textbf{4}, lead to similar conclusions. For the larger sample size, both the Cox model and the frailty model exhibit conservative type I error, especially in the shorter observation window setting. Overall, the AUMCF is the only method that consistently controls type I error and provides accurate estimation and inference, without a major loss in power compared to parametric alternatives and with improved power over win ratio methods.  

With respect to the proposed covariate adjustment method, \textbf{Web Figure 1} presents the relative efficiency of the adjusted and unadjusted estimators for scenarios (i) and (ii) under the null using both an informative and uninformative covariate. As expected, augmentation with an uninformative covariate provides no benefit, whereas augmentation with an informative covariate improves efficiency, especially for the independent competing risks scenario where the covariate imbalance is more significant. \textbf{Web Figures 2} and \textbf{3} correspond to the smaller and larger sample sizes, respectively, and show similar results.

\begin{table}[htbp]
\centering

\textbf{A: Null Setting} \\

\vspace{0.5em}

\scalebox{0.9}{
\begin{tabular}{lrrrr|rrrr}
\toprule
& \multicolumn{4}{c|}{$\tau = 1$} & \multicolumn{4}{c}{$\tau = 4$} \\
\cmidrule(lr){2-5} \cmidrule(lr){6-9}
\textbf{ICR Model} & Bias & ESE & ASE & Type I Error & Bias & ESE & ASE & Type I Error \\
\midrule
AUMCF        & 0.0002 & 0.058 & 0.058 & 0.050 & 0.0098 & 0.530 & 0.527 & 0.052 \\
CoxPH       & 0.0089 & 0.127 & 0.125 & 0.052 & 0.0077 & 0.111 & 0.109 & 0.049 \\
Frailty      & 0.0125 & 0.113 & 0.113 & 0.047 & 0.0046 & 0.072 & 0.072 & 0.047 \\
LWYY         & 0.0072 & 0.111 & 0.110 & 0.051 & 0.0037 & 0.072 & 0.071 & 0.053 \\
NB           & 0.0072 & 0.111 & 0.112 & 0.047 & 0.0037 & 0.072 & 0.072 & 0.049 \\
LWR     & 0.0070 & 0.132 & 0.131 & 0.051 & 0.0064 & 0.117 & 0.116 & 0.050 \\
SWR    & 0.0076 & 0.134 & 0.132 & 0.054 & 0.0065 & 0.118 & 0.118 & 0.048 \\
\midrule
\textbf{Frailty Model} & \multicolumn{4}{c|}{} & \multicolumn{4}{c}{} \\
AUMCF        & -0.0004 & 0.087 & 0.085 & 0.051 & -0.0063 & 0.777 & 0.772 & 0.050 \\
CoxPH       & 0.0141  & 0.169 & 0.164 & 0.050 & 0.0105 & 0.144 & 0.141 & 0.048 \\
Frailty      & 0.0522  & 0.259 & 0.235 & 0.057 & 0.0384 & 0.246 & 0.232 & 0.050 \\
LWYY         & 0.0230  & 0.219 & 0.208 & 0.054 & 0.0169 & 0.196 & 0.188 & 0.053 \\
NB           & 0.0260  & 0.262 & 0.217 & 0.058 & 0.0207 & 0.227 & 0.205 & 0.066 \\
LWR     & 0.0136  & 0.171 & 0.169 & 0.048 & 0.0114 & 0.149 & 0.148 & 0.051 \\
SWR    & 0.0137  & 0.171 & 0.169 & 0.049 & 0.0112 & 0.150 & 0.148 & 0.052 \\
\bottomrule
\end{tabular}
}
\vspace{1em}

\textbf{B: Non-null Setting} \\

\vspace{0.5em}
\scalebox{0.9}{
\begin{tabular}{lrrrrr|rrrrr}
\toprule
& \multicolumn{5}{c|}{$\tau = 1$} & \multicolumn{5}{c}{$\tau = 4$} \\
\cmidrule(lr){2-6} \cmidrule(lr){7-11}
\textbf{ICR Model} & Bias & ESE & ASE & Power & CP & Bias & ESE & ASE & Power & CP \\
\midrule
AUMCF        & 0.0006 & 0.065 & 0.064 & 0.826 & 0.949 & 0.0035 & 0.602 & 0.593 & 0.987 & 0.946 \\
CoxPH       & 0.0058 & 0.091 & 0.121 & 0.664 & 0.952 & 0.0034 & 0.082 & 0.109 & 0.759 & 0.948 \\
Frailty      & 0.0061 & 0.074 & 0.104 & 0.885 & 0.953 & -0.0018 & 0.049 & 0.067 & 0.995 & 0.948 \\
LWYY         & 0.0044 & 0.073 & 0.102 & 0.912 & 0.950 & 0.0019 & 0.048 & 0.065 & 0.999 & 0.945 \\
NB           & 0.0044 & 0.073 & 0.104 & 0.907 & 0.952 & 0.0019 & 0.048 & 0.067 & 0.999 & 0.951 \\
LWR     & 0.0097 & 0.169 & 0.169 & 0.589 & - & 0.0075 & 0.146 & 0.145 & 0.511 & -\\
SWR    & 0.0102 & 0.163 & 0.162 & 0.425 & - & 0.0074 & 0.137 & 0.136 & 0.256 & - \\
\midrule
\textbf{Frailty Model} & \multicolumn{5}{c|}{} & \multicolumn{5}{c}{} \\
AUMCF        & 0.0003 & 0.100 & 0.100 & 0.362 & 0.949 & -0.0002 & 0.922 & 0.918 & 0.455 & 0.949 \\
CoxPH       & 0.0080 & 0.140 & 0.160 & 0.161 & 0.948 & -0.0013 & 0.124 & 0.138 & 0.148 & 0.948 \\
Frailty      & 0.0331 & 0.176 & 0.229 & 0.310 & 0.943 & 0.0257 & 0.171 & 0.226 & 0.300 & 0.948 \\
LWYY         & 0.0150 & 0.151 & 0.202 & 0.386 & 0.946 & 0.0122 & 0.136 & 0.185 & 0.451 & 0.947 \\
NB           & 0.0178 & 0.212 & 0.213 & 0.392 & 0.943 & 0.0207 & 0.178 & 0.202 & 0.453 & 0.937 \\
LWR     & 0.0170 & 0.193 & 0.190 & 0.147 & - & 0.0120 & 0.164 & 0.162 & 0.117 & - \\
SWR    & 0.0169 & 0.192 & 0.189 & 0.134 & - & 0.0117 & 0.163 & 0.162 & 0.107 & - \\
\bottomrule
\end{tabular}
}

\caption{Operating characteristics of the AUMCF estimator and the comparison methods with treatment arm sample sizes of $n_1 = n_2 = 200$  under the independent competing risks (ICR) and frailty models for observation periods of $\tau = 1, 4$. The comparison methods include the Cox Proportional Hazards Model (CoxPH), the shared frailty model (Frailty), the LWYY method (LWYY), negative binomial regression (NB), the standard win ratio (SWR), and the last-event-assisted win ratio (LWR). ESE is the empirical standard error across simulation replications. ASE is the mean of the asymptotic standard errors across replicates. CP is the empirical coverage probability of the 95\% confidence interval. Results are shown for (A) type I error in the null setting corresponding to no difference in the AUMCFs across treatment arms and (B) power in a non-null setting where the difference in the AUMCFs across treatment arms is non-zero. }
\label{tab:c1ac2_n200}
\end{table}

\begin{table}[htbp]
\centering

\textbf{A: Null Setting} \\

\vspace{0.5em}
\scalebox{0.9}{
\begin{tabular}{lrrrr|rrrr}
\toprule
& \multicolumn{4}{c|}{$\tau = 4$} & \multicolumn{4}{c}{$\tau = 6$} \\
\cmidrule(lr){2-5} \cmidrule(lr){6-9}
Method & Bias & ESE & ASE & Type I Error & Bias & ESE & ASE & Type I Error \\
\midrule
AUMCF    & 0.0014 & 0.443 & 0.440 & 0.052 & -0.0068 & 0.814 & 0.802 & 0.054 \\
CoxPH   & 0.0070 & 0.113 & 0.111 & 0.051 & 0.0080  & 0.112 & 0.110 & 0.052 \\
Frailty  & 0.0067 & 0.086 & 0.086 & 0.046 & 0.0051  & 0.082 & 0.082 & 0.049 \\
LWYY     & 0.0042 & 0.086 & 0.084 & 0.049 & 0.0038  & 0.083 & 0.080 & 0.055 \\
NB       & 0.0044 & 0.087 & 0.087 & 0.048 & 0.0043  & 0.085 & 0.083 & 0.052 \\
LWR & 0.0057 & 0.117 & 0.117 & 0.050 & 0.0034  & 0.116 & 0.116 & 0.049 \\
SWR& 0.0060 & 0.119 & 0.119 & 0.050 & 0.0034  & 0.118 & 0.118 & 0.049 \\
\bottomrule
\end{tabular}
}
\vspace{1em}

\textbf{B: Non-null Setting} \\

\vspace{0.5em}
\scalebox{0.9}{
\begin{tabular}{lrrrrr|rrrrr}
\toprule
& \multicolumn{5}{c|}{$\tau = 4$} & \multicolumn{5}{c}{$\tau = 6$} \\
\cmidrule(lr){2-6} \cmidrule(lr){7-11}
Method & Bias & ESE & ASE & Power & CP & Bias & ESE & ASE & Power & CP \\
\midrule
AUMCF    & 0.0092 & 0.487 & 0.486 & 0.879 & 0.948 & 0.0069 & 0.956 & 0.942 & 0.980 & 0.946 \\
CoxPH   & 0.0347 & 0.099 & 0.110 & 0.181 & 0.939 & 0.0373 & 0.099 & 0.110 & 0.189 & 0.936 \\
Frailty  & 0.0393 & 0.060 & 0.080 & 0.961 & 0.900 & 0.0417 & 0.057 & 0.076 & 0.984 & 0.883 \\
LWYY     & 0.0456 & 0.056 & 0.078 & 0.994 & 0.870 & 0.0549 & 0.052 & 0.074 & 0.999 & 0.794 \\
NB       & 0.0466 & 0.056 & 0.079 & 0.991 & 0.872 & 0.0563 & 0.053 & 0.075 & 0.999 & 0.792 \\
LWR & 0.0040 & 0.130 & 0.130 & 0.160 & - & 0.0039 & 0.129 & 0.128 & 0.138 & - \\
SWR& 0.0038 & 0.120 & 0.120 & 0.050 & - & 0.0032 & 0.119 & 0.119 & 0.049 & - \\
\bottomrule
\end{tabular}
}
\caption{Operating characteristics of the AUMCF estimator and the comparison methods with treatment arm sample sizes of $n_1 = n_2 = 200$  under the time-varying treatment effect model for observation periods of $\tau = 4, 6$. The comparison methods include the Cox Proportional Hazards Model (CoxPH), the shared frailty model (Frailty), the LWYY method (LWYY), negative binomial regression (NB), the standard win ratio (SWR), and the last-event-assisted win ratio (LWR). ESE is the empirical standard error across simulation replications. ASE is the mean of the asymptotic standard errors across replicates. CP is the empirical coverage probability of the 95\% confidence interval. Results are shown for (A) type I error in the null setting corresponding to no difference in the AUMCFs across treatment arms and (B) power in a non-null setting where the difference in the AUMCFs across treatment arms is non-zero. }
\label{tab:c3_n200}
\end{table}

% -------------------------------------------------

% -------------------------------------------------
\vspace{-12mm}
\section{Discussion}
\label{s:discuss}
We have rigorously defined and evaluated an estimand for quantifying treatment efficacy in comparative clinical studies with a multiple event endpoint: the area under the mean cumulative function (AUMCF). The AUMCF is a natural generalization of the restricted mean time lost to the multiple event-time setting. It is applicable to both recurrences of a single type of event and to events of multiple types. Clinically, the AUMCF is the expected total event-free time lost due to all event occurrences, which quantifies the average patient's cumulative disease burden across time. We demonstrated how to contrast the efficacy of two treatments with a nonparametric estimation and inference procedure for the difference in AUMCFs. Moreover, terminal events are accounted for by stopping the counting process, allowing for assumption-free handling of competing risks.

Our approach is not without limitations. First, as presented, the AUMCF weights all events equally in assessing the treatment effect, regardless of whether they are severe or minor. A simple way to accommodate events with differing levels of severity is to consider multiple event processes and obtain a weighted sum of the corresponding AUMCFs. In particular, one can define $K$ potentially-dependent counting processes, $N^*_{k}(t)$ for $k = 1, \dots, K$, for each of the events of interest. For each counting process, the MCF is defined as $m_{k}(t) = \Ebb\{N^*_{k}(t)\}$ while the corresponding AUMCF is $\theta_{k} = \int_{0}^{\tau}m_{k}(t)dt$. The total disease burden can be summarized with a weighted average of the per-process AUMCFs as $\theta_{W} = \sum_{k=1}^K w_k\theta_{k}$, where $w_k > 0$ is a weight reflecting the relative severity or importance of the $k$th type of event. Given a set of weights, $\theta_{W}$ can be estimated by $\widehat{\theta}_{W} =\sum_{k=1}^K w_k \widehat{\theta}_k$. 

Second, the proposed augmentation estimator requires independence of the baseline covariates and the censoring time. While this assumption is reasonable in the setting of randomized controlled trials with administrative censoring, it would be restrictive in observational studies. Future work will extend AUMCF analysis to observational data, including the possibility of relaxing the independence assumption. Lastly, while we have introduced an augmented estimator to adjust for baseline covariates, regression modeling to characterize the relationship between individual covariates and the AUMCF may be desirable, particularly in observational settings. 

%  The \backmatter command formats the subsequent headings so that they
%  are in the journal style. Please keep this command in your document
%  in this position, right after the final section of the main part of 
%  the paper and right before the Acknowledgements, Supplementary Materials,
%  and References sections. 

\backmatter

%  This section is optional. Here is where you will want to cite
%  grants, people who helped with the paper, etc. But keep it short!

\vspace{-10mm}
\section*{Acknowledgments}

J. Gronsbell is grateful for support of an NSERC Discovery Grant (RGPIN-2021-03734). J. Gronsbell, Z. McCaw, and I. Nogues contributed equally.  \vspace*{-8pt}

\vspace{-6mm}
\section*{Supplementary Material}
Web Appendices, Tables, and Figures referenced in Sections 2 and 3 and code are available with this paper at the Biometrics website on Oxford Academic. The \texttt{R} package, \texttt{MCC}, developed for the proposed method is available at \url{https://github.com/zrmacc/MCC}.

\section*{Data Availability}
The BEST data in this paper were provided by permission and cannot be shared publicly. 

\vspace{-10mm}
\bibliographystyle{biom}
\bibliography{biomsample}

%\appendix

%  To get the journal style of heading for an appendix, mimic the following.

%\section{}
%\subsection{Title of appendix}

%Put your short appendix here. Remember, longer appendices are
%possible when presented as Supplementary Web Material. Please 
%review and follow the journal policy for this material, available
%under Instructions for Authors at \texttt{http://www.biometrics.tibs.org}.

\label{lastpage}

\end{document}

\include{GrandMacros}

\makeatletter
\renewcommand{\fnum@figure}{Web \figurename~\thefigure}
\makeatother

\makeatletter
\renewcommand{\fnum@table}{Web \tablename~\thetable}
\makeatother

\renewcommand{\thesection}{\Alph{section}}
\titleformat{\section}[hang]
  {\normalfont\Large\bfseries} % Format for section title
  {Web Appendix \thesection:}  % Prefix
  {0.25em} % Spacing between prefix and title
  {}
    
\maketitle
\clearpage
\section{Asymptotic properties of the AUMCF}

\subsection{One-sample estimator}
{\blue{Here we derive the influence function of our proposed estimator of the AUMCF, which is the basis of Theorem 1 in the main text.}}
\begin{proof}
{\blue{Recall the estimator of the AUMCF in the $j^{th}$ treatment arm and its population-level counterpart, defined respectively as 
\begin{align}
\widehat{\theta}_{j}=\int_0^\tau (\tau-u)\widehat{S}_{_j}^{D}(u)\frac{d \bar{N}_j(u)}{\bar{Y}_j(u)} \quad \mbox{and} \quad \theta_{j} =  \int_0^\tau (\tau-u)S_{_j}^{D}(u)dR_{j}(u). \label{eq: defn}
\end{align}
Throughout, we assume that $P(X \ge \tau) > 0$. \\}}

{\blue{Using the definitions in (\ref{eq: defn}) and letting $d\widehat{R}_j(u) = \frac{d \bar{N}_j(u)}{\bar{Y}_j(u)}$, we have that 
\begin{align}\label{eq: main-kj}
n_j^{1/2}(\widehat{\theta}_{j}-\theta_{j}) &=  n_j^{1/2} \int_0^\tau (\tau-u)\left\{\widehat{S}_{_j}^{D}(u)-S_{_j}^{D}(u)\right\}dR_{j}(u) \nonumber \\ 
& +  n_j^{1/2} \int_0^\tau (\tau-u)\widehat{S}_{_j}^{D}(u) d\left\{\widehat{R}_j(u) - R_j(u)\right\}. 
\end{align}
We begin by considering the first term in (\ref{eq: main-kj}), namely $n_j^{1/2} \int_0^\tau (\tau-u)\left\{\widehat{S}_{_j}^{D}(u)-S_{_j}^{D}(u)\right\}dR_{j}(u)$.  Let $\widehat{\Lambda}_{_j}^{D}(\cdot)$ be the Nelson-Aalen estimator of the cumulative hazard function of the terminal event, denoted as $\Lambda_{_j}^{D}(\cdot)$.  Since $\widehat{S}_{_j}^{D}(t)$ is asymptotically equivalent to $e^{-\widehat{\Lambda}_{_j}^{D}(t)}$ and $S_{_j}^{D}(t) = e^{-\Lambda_{_j}^{D}(t)}$, it follows that
\begin{align}
&n_j^{1/2}\int_0^\tau (\tau-u)\left\{\widehat{S}_{_j}^{D}(u)-S_{_j}^{D}(u)\right\}dR_{j}(u) =  -n_j^{1/2}\int_0^\tau (\tau-u) S_{_j}^{D}(u) \left\{1 - \frac{\widehat{S}_{_j}^{D}(u)}{S_{_j}^{D}(u)}\right\}dR_{j}(u) \nonumber \\
&=  - n_j^{1/2}\int_0^\tau (\tau-u) S_{_j}^{D}(u) \left\{1 - e^{- \left\{ \widehat{\Lambda}_{_j}^{D}(u) - \Lambda_{_j}^{D}(u) \right\} } \right\} dR_{j}(u)  +  o_p(1) \label{first-term: simp-1}.
\end{align}
By a first-order Taylor expansion, the term $1 -e^{- \left\{ \widehat{\Lambda}_{_j}^{D}(t) - \Lambda_{_j}^{D}(t) \right\} }$ is asymptotically equivalent to $\widehat{\Lambda}_{_j}^{D}(t) - \Lambda_{_j}^{D}(t)$. Using this equivalence and recalling that $m_j(t) = \int_0^t S_{_j}^{D}(u) dR_j(u)$, we can further simplify (\ref{first-term: simp-1}) as
\begin{align}
&n_j^{1/2}\int_0^\tau (\tau-u)\left\{\widehat{S}_{_j}^{D}(u)-S_{_j}^{D}(u)\right\}dR_{j}(u) \nonumber  \\
&=  -n_j^{1/2}\int_0^\tau (\tau-u) \left\{\widehat{\Lambda}_{_j}^{D}(u) - \Lambda_{_j}^{D}(u) \right\} dm_{j}(u) + o_p(1).\label{first-term: simp-2} 
%&= -n_j^{1/2}\int_0^\tau (\tau-u)  \left\{\widehat{\Lambda}_{_j}^{D}(u) - \Lambda_{_j}^{D}(u) \right\} dm_j(u) + o_p(1)  \label{first-term: simp-2} 
\end{align}
Next, define $M_{ij}^D(t) = N_{ij}^D(t) - \int_0^t Y_{ij}(u) d\Lambda_j^D(u)$ where $N_{ij}^D(t) = \mathbb{I}( X_{ij} \le t, \delta_{ij} = 1)$, and $Y_{ij}(t) = \mathbb{I}( X_{ij} \ge t)$.  Then, for $t \le  \max_i X_i$,
\begin{align*}
n_j^{1/2}\left\{\widehat{\Lambda}_{_j}^{D}(t) - \Lambda_{_j}^{D}(t) \right\}  &= n_j^{-1/2} \sum_{i=1}^{n_j} \int_0^t  \frac{dM_{ij}^{D}(u)}{\bar{Y}_j(u)/n_j}  %\\
%&= n_j^{-1/2} \sum_{i=1}^{n_j} \int_0^\tau \mathbb{I}(u \le t)  \frac{dM_{ij}^{D}(u)}{\bar{Y}_j(u)/n_j}. 
\end{align*}
Since  $\bar{Y}_j(t)/n_j \overset{p}{\to} \pi_j(t)$ where $\pi_j(t) = P( X_j \ge t)$, it follows from the Martingale Central Limit Theorem (\cite{fleming2013counting}, Theorem 4.2.1) that
\begin{align}\label{lambdaD-simp}
n_j^{1/2}\left\{\widehat{\Lambda}_{_j}^{D}(t) - \Lambda_{_j}^{D}(t) \right\} &= n_j^{-1/2} \sum_{i=1}^{n_j} \int_0^t   \frac{dM_{ij}^{D}(u)}{\pi_j(u)} + o_p(1) 
\end{align}
for $t \in [0, \tau]$.  Combining the results in  (\ref{first-term: simp-2}) and (\ref{lambdaD-simp}), 
%Using integration by parts to re-express (\ref{first-term: simp-2}) and applying the result in (\ref{lambdaD-simp}), 
\begin{align}
& n_j^{1/2}\int_0^\tau (\tau-u)\left\{\widehat{S}_{_j}^{D}(u)-S_{_j}^{D}(u)\right\}dR_{j}(u) \nonumber \\
&=  -  \int_0^\tau (\tau-v) \left\{  n_j^{-1/2} \sum_{i=1}^{n_j} \int_0^v   \frac{dM_{ij}^{D}(u)}{\pi_j(u)}\right\} dm_{j}(v) + o_p(1) \nonumber\\
&= - n_j^{-1/2}  \sum_{i=1}^{n_j} \int_0^\tau  \left\{ \int_u^\tau  (\tau-v)  dm_{j}(v) \right\}  \frac{dM_{ij}^{D}(u)}{\pi_j(u)}  + o_p(1) \label{eq: first-term}
%&= - n_j^{-1/2}  \sum_{i=1}^{n_j} \int_0^\tau  \left\{ \int_u^\tau  (\tau-v)  dm_j(v) \right\} \frac{dM_{ij}^{D}(u)}{\pi_j(u)}  + o_p(1)  \label{eq: first-term}
%&= -n_j^{1/2} \left[ \int_0^\tau m_j(u) \left\{ \widehat{\Lambda}_{_j}^{D}(u) - \Lambda_{_j}^{D}(u) \right\} \, du  - \int_0^\tau m_j(u) (\tau - u) \, d\left\{ \widehat{\Lambda}_{_j}^{D}(u) - \Lambda_{_j}^{D}(u) \right\} \right] +o_p(1) \\
%&=  -n_j^{-1/2} \sum_{i=1}^{n_j} \left[ \int_0^\tau m_j(u)  \frac{dM_{ij}^{D}(u)}{\pi_j(u)} - \int_0^{\tau}  m_j(u)(\tau-u)  \frac{dM_{ij}^{D}(u)}{\pi_j(u)} \right] + o_p(1)
\end{align}
Returning to the second term of (\ref{eq: main-kj}), namely  $n_j^{1/2} \int_0^\tau (\tau-u)\widehat{S}_{_j}^{D}(u) d\left\{\widehat{R}_j(u) - R_j(u)\right\}$, the consistency of $\widehat{S}_{_j}^{D}(t)$ for $S_{_j}^{D}(t)$ implies that
\begin{align}
&  n_j^{1/2} \int_0^\tau (\tau-u)\widehat{S}_{_j}^{D}(u) d\left\{\widehat{R}_j(u) - R_j(u) \right\} =   n_j^{1/2} \int_0^\tau (\tau-u){S}_{_j}^{D}(u) d\left\{\widehat{R}_j(u) - R_j(u) \right\} + o_p(1) \label{eq: scd-term}.
\end{align}
Letting $M_{ij}(t) =  N_{ij}(t) - \int_{0}^t Y_{ij}(u) dR_j(u)$, it follows from \cite{ghosh2000nonparametric} that for $t \le \max_i X_i$,
\begin{align*}
 n_j^{1/2} \left\{ \widehat{R}_j(t) - R_j(t)\right\}  = n_j^{-1/2} \sum_{i=1}^{n_j} \int_0^t  \frac{dM_{ij}(u)}{\bar{Y}_j(u)/n_j}. 
\end{align*}
Although the $M_{ij}(\cdot)$ are not martingales and we cannot apply the Martingale Central Limit Theorem, arguments from empirical process theory (e.g., \cite{ghosh2000nonparametric, lin2000semiparametric}) yield
\begin{align*}
 n_j^{1/2} \left\{ \widehat{R}_j(t) - R_j(t) \right\}  = n_j^{-1/2} \sum_{i=1}^{n_j} \int_0^\tau \mathbb{I}(u \le t)  \frac{dM_{ij}(u)}{\pi_j(u)} +o_p(1) 
\end{align*}
for $t \in [0, \tau]$.  This result then implies that the term in (\ref{eq: scd-term}) can be further simplified as
\begin{align}\label{second-term: simp-1}
& \int_0^\tau (\tau-u)\widehat{S}_{_j}^{D}(u) d\left\{\widehat{R}_j(u) - R_j(u) \right\} =  n_j^{-1/2} \sum_{i=1}^{n_j}\int_0^\tau (\tau-u)S_{_j}^{D}(u)\frac{dM_{ij}(u)}{\pi_j(u)} + o_p(1).
\end{align}
Putting together (\ref{eq: first-term}) and (\ref{second-term: simp-1}), $n_j^{1/2}(\widehat{\theta}_{j}-\theta_{j})$ is asymptotically equivalent to
\begin{align*}
n_j^{1/2}(\widehat{\theta}_{j}-\theta_{j}) =  n_j^{-1/2} \sum_{i=1}^{n_j}\Psi_{ij} +o_p(1)
\end{align*}
where $\Psi_{ij} =  \int_0^\tau (\tau-u)S_{_j}^{D}(u)\frac{dM_{ij}(u)}{\pi_j(u)} - \int_0^\tau  \left\{ \int_u^\tau  (\tau-v)  dm_j(v) \right\}  \frac{dM_{ij}^{D}(u)}{\pi_j(u)}.$  It then follows that $n_j^{1/2}(\widehat{\theta}_{j}-\theta_{j})$ converges weakly to a zero-mean normal distribution with variance $\Sigma_j = \mathbb{E}( \Psi_{ij}^2)$.
}}
\end{proof}

\subsection{Two-sample estimator}
{\blue{We next verify Theorem 1 stated in main text, which provides the asymptotic distribution of the difference in the AUMCF estimators across two treatment arms.}}

\begin{theorem}\label{thm: delta-IF}
 {\blue{Let $n = n_1 + n_2$ and assume that $ \lim_{n \to \infty} n_j /n = \rho_j  \in (0,1)$ for $j = 1,2$.  Then, $\left(\dfrac{n_1 n_2}{n}\right)^{1/2}( \hat{\Delta} - \Delta)$ converges weakly to a mean-zero normal distribution with variance $\Sigma_\Delta = \rho_2 \Sigma_1 + \rho_1\Sigma_2$ 
 where $\Sigma_j = \mathbb{E}( \Psi_{ij}^2)$ and  $\Psi_{ij} =  \int_0^\tau (\tau-u)S_{_j}^{D}(u)\frac{dM_{ij}(u)}{\pi_j(u)} - \int_0^\tau  \left\{ \int_u^\tau  (\tau-v)  dm_j(v) \right\}  \frac{dM_{ij}^{D}(u)}{\pi_j(u)}.$
}}
\end{theorem}

\begin{proof}
{\blue{In Web Appendix A.1, we showed that
\begin{align*}
n_j^{1/2}(\widehat{\theta}_{j}-\theta_{j}) =  n_j^{-1/2} \sum_{i=1}^{n_j}\Psi_{ij} +o_p(1)
\end{align*}
where $\Psi_{ij} =  \int_0^\tau (\tau-u)S_{_j}^{D}(u)\frac{dM_{ij}(u)}{\pi_j(u)} - \int_0^\tau  \left\{ \int_u^\tau  (\tau-v)  dm_j(v) \right\}  \frac{dM_{ij}^{D}(u)}{\pi_j(u)}$ for $j = 1,2$.  Recall that 
$$\left(\dfrac{n_1 n_2}{n}\right)^{1/2} ( \hat{\Delta} - \Delta) = \left(\dfrac{n_1 n_2}{n}\right)^{1/2}\{ (\widehat{\theta}_{1}- \theta_1) - (\widehat{\theta}_{2}- \theta_2)\}.$$  
Therefore, 
\begin{align*}
 \left(\dfrac{n_1 n_2}{n}\right)^{1/2}( \hat{\Delta} - \Delta) &=   \left(\dfrac{n_2}{n}\right)^{1/2} \left( \frac{1}{\sqrt{n_1}}  \right) \sum_{i=1}^{n_1}\Psi_{i1} -  \left(\dfrac{n_1}{n}\right)^{1/2} \left( \frac{1}{\sqrt{n_2}} \right)\sum_{i=1}^{n_2}\Psi_{i2} + o_p(1) \\
  &= \left(\frac{\rho_2}{n_1}\right)^{1/2}  \sum_{i=1}^{n_1}  \Psi_{i1} -  \left(\frac{\rho_1}{n_2}\right)^{1/2}  \sum_{i=1}^{n_2}  \Psi_{i2} + o_p(1)
\end{align*}
It then follows that $\left(\dfrac{n_1 n_2}{n}\right)^{1/2}( \hat{\Delta} - \Delta)$ converges weakly to a mean-zero normal distribution with variance given by $\Sigma_\Delta = \rho_2\Sigma_1 + \rho_1\Sigma_2$.}}    
\end{proof}

\subsection{Inference for the two-sample estimator}
{\blue{To perform inference on $\widehat{\Delta}$, note that $\Sigma_\Delta$ can be consistently estimated by 
$\widehat{\Sigma}_\Delta =  \frac{n_2}{n}\widehat{\Sigma}_{1} + \frac{n_1}{n}\widehat{\Sigma}_2$
where $\Sigma_j = n_j^{-1} \sum_{i =1 }^{n_j} \widehat{\Psi}_{ij}^2$ and $\widehat{\Psi}_{ij}$ is obtained by replacing the unknown quantities in $\Psi_{ij}$ with their respective estimators \citep{ghosh2000nonparametric}.  That is, $$\widehat{\Sigma}_j=   \frac{1}{n_j}\sum_{i=1}^{n_j} \left( \int_0^\tau \left[  (\tau-u)\widehat{S}_{_j}^{D}(u) \frac{ d\widehat{M}_{ij}(u)}{n_j^{-1}\bar{Y}_j(u)} - \left\{ \int_u^\tau(\tau-v)d\widehat{m}_{j}(v)\right\}\frac{d\widehat{M}_{ij}^{D}(u)}{n_j^{-1}\bar{Y}_j(u)}\right] \right)^2$$
where the various estimators are defined as
$$\widehat{m}_{j}(t)=\int_0^t \widehat{S}_{_j}^{D}(u)\frac{d \bar{N}_j(u)}{\bar{Y}_j(u)},$$
$$\widehat{M}_{ij}(t)=N_{ij}(t)-\int_0^t \mathbb{I}(X_{ij}\ge u) \frac{d\bar{N}_j(u)}{\bar{Y}_j(u)},$$
$$ \widehat{M}_{ij}^{D}(t)=\mathbb{I}(X_{ij}\le t)\delta_{ij}-\int_0^t \mathbb{I}(X_{ij}\ge u) \frac{d\bar{N}_j^{D}(u)}{\bar{Y}_j(u)},$$
and $\bar{N}_j^{D}(t)=\sum_{i=1}^{n_j} \mathbb{I}(X_{ij} \le t)\delta_{ij}.$ 
 A corresponding $100(1-\alpha)$\% confidence interval for $\Delta$ can  be constructed as $\left[\widehat{\Delta}- z_{1-\alpha/2} \frac{\sqrt{\widehat{\Sigma}_{\Delta}}}{\sqrt{(n_1n_2)/n}},  \widehat{\Delta}+ z_{1-\alpha/2}  \frac{ \sqrt{\widehat{\Sigma}}_{\Delta}}{\sqrt{(n_1n_2)/n}}\right]$ where $z_{1-\alpha/2}$ is the $(1-\alpha/2)$ quantile of the standard Normal distribution. To test the null hypothesis of $H_0: \Delta=0$, we can use the fact that under the null,
$$\left( \frac{n_1 n_2}{n}\right)^{1/2}\frac{\widehat{\Delta}}{\sqrt{\widehat{\Sigma}}_{\Delta}} \sim^{approx.} N(0, 1).$$
The $p$-value for the two-sided test of $H_1: \Delta \ne 0$ can then be calculated as 
$$\mathbb{P}\left(|Z|\ge \sqrt{(n_1n_2)/n}\frac{|\widehat{\Delta}|}{\sqrt{\widehat{\Sigma}}_{\Delta}}\right)$$\\
for $Z \sim N(0,1)$. 
}}

\subsection{Covariate-adjusted estimator}
{\blue{Next we verify Theorem 2 in the main text, which summarizes the asymptotic properties of the proposed augmented estimator of the AUMCF in the two-sample setting.}}

\begin{theorem}\label{thm: cov-adj}
{\blue{Let $\bm{\beta}_0 = \Sigma_{\bm{W}}^{-1} {\bm \gamma} _0$ where $\Sigma_{\bm{W}} =  \rho_2 \text{Var}(\bm{W}_{i1}) +  \rho_1 \text{Var}(\bm{W}_{i2})$ and ${\bm \gamma} _0 = \rho_2  \text{Cov}(\bm{W}_{i1}, \Psi_{i1}) +  \rho_1  \text{Cov}(\bm{W}_{i2}, \Psi_{i2})$ and assume that $ \lim_{n \to \infty} n_j /n = \rho_j  \in (0,1)$ for $j = 1, 2$.   Then, $\hat{\Delta}^{\aug}$ is consistent for $\Delta$ and $\left(\dfrac{n_1 n_2}{n}\right)^{1/2}( \hat{\Delta}^{\aug} - \Delta)$ converges weakly to a mean-zero normal distribution with variance $\Sigma_\Delta - {\bm \gamma}_0^T  {\Sigma}_{\bm{W}}^{-1}  {\bm \gamma}_0$. }}   
\end{theorem}

\begin{proof}
{\blue{To show that $\widehat{\Delta}^{\aug}$ is consistent for $\Delta$, we first consider the consistency of $\widehat{\bm \beta}=\widehat{\Sigma}_{\bm{W}}^{-1}\widehat{\bm \gamma}$ for $\bm{\beta}_0$. Recall that
\begin{align*}
    \widehat{\bm \gamma}= \left(\frac{n_2}{n}\right) \frac{1}{n_1} \sum_{i=1}^{n_1}  \left(\bm{W}_{i1}-\overline{\bm W}_1\right) \widehat{\Psi}_{i1} + \left(\frac{n_1}{n}\right) \frac{1}{n_2} \sum_{i=1}^{n_2}  \left(\bm{W}_{i2}-\overline{\bm W}_2\right) \widehat{\Psi}_{i2}
\end{align*}
and
\begin{align*}
    \widehat{\Sigma}_{\bm{W}}= \left(\frac{n_2}{n}\right) \frac{1}{n_1 }\sum_{i=1}^{n_1} \left(\bm{W}_{i1}-\overline{\bm W}_1\right) \left(\bm{W}_{i1}-\overline{\bm W}_1\right)^T + \left(\frac{n_1}{n}\right) \frac{1}{n_2 }\sum_{i=1}^{n_2} \left(\bm{W}_{i2}-\overline{\bm W}_2\right) \left(\bm{W}_{i2}-\overline{\bm W}_2\right)^T
\end{align*}
Due to the consistency of $\widehat{\Psi}_{ij}$ for ${\Psi}_{ij}$ and the fact that $\lim_{n \to \infty} n_j /n = \rho_j$, $\widehat{\bm \gamma}$ is consistent for $\bm{\gamma}_0$.  Similarly, $\widehat{\Sigma}_{\bm{W}}$ is consistent for $\Sigma_{\bm{W}}$ and hence $\widehat{\bm \beta}$ is consistent for $\bm{\beta}_0$.  This result, together with the facts that
 $\widehat{\Delta} \overset{p}\to \Delta$ and $\overline{\bm{W}}_1 - \overline{\bm{W}}_2 \overset{p}\to 0$ due to randomization, imply the consistency of $\widehat{\Delta}^{aug}$ for $\Delta$.\\
}}

{\blue{
For the asymptotic normality of $\widehat{\Delta}^{\aug}$, note that the consistency of $\widehat{\bm \beta}$ for ${\bm \beta}_0$ implies that
$$\left(\dfrac{n_1 n_2}{n}\right)^{1/2}( \hat{\Delta}^{\aug} - \Delta) = \left(\dfrac{n_1 n_2}{n}\right)^{1/2}\{(\widehat{\Delta}  -  \Delta) -  {\bm \beta}_0^T (\overline{\bm{W}}_1 - \overline{\bm{W}}_2)\} + o_p(1).$$
From Theorem \ref{thm: delta-IF}, it follows that
\begin{align}
 \left(\dfrac{n_1 n_2}{n}\right)^{1/2}( \hat{\Delta}^{\aug} - \Delta)  &= \left(\frac{\rho_2}{n_1}\right)^{1/2}  \sum_{i=1}^{n_1}  \Psi_{i1} -  \left(\frac{\rho_1}{n_2}\right)^{1/2}  \sum_{i=1}^{n_2}  \Psi_{i2} - \\ & {\bm \gamma}_0^T  \Sigma_{\bm{W}}^{-1} \left( \left(\frac{\rho_2}{n_1}\right)^{1/2}  \sum_{i=1}^{n_1}  \bm{W}_{i1} -  \left(\frac{\rho_1}{n_2}\right)^{1/2}  \sum_{i=1}^{n_2}  \bm{W}_{i2}   \right)+ o_p(1) \label{eq:IF-aug}
\end{align}
Note that the mean of the influence function in (\ref{eq:IF-aug}) is 0 as treatment is randomized.  By the Central Limit Theorem, $\left(\dfrac{n_1 n_2}{n}\right)^{1/2}( \hat{\Delta}^{\aug} - \Delta)$ converges weakly to a mean-zero normal distribution with variance given by 
\begin{align*}
\Sigma_\Delta^{\aug} &=  \rho_2 \text{Var}\left( \Psi_{i1} \right) +  \rho_1 \text{Var}\left( \Psi_{i2} \right) +  {\bm \gamma}_0^T  \Sigma_{\bm{W}}^{-1}  \left\{  \rho_2 \text{Var}\left( \bm{W}_{i1} \right) +  \rho_1 \text{Var}\left( \bm{W}_{i2} \right)  \right\} \Sigma_{\bm{W}}^{-1}{\bm \gamma}_0 - \\
&  2 \left\{ \rho_2 \text{Cov}\left( \Psi_{i1}, \bm{W}_{i1} \right)  +  \rho_1 \text{Cov}\left( \Psi_{i2}, \bm{W}_{i2} \right)  \right\}    \Sigma_{\bm{W}}^{-1}{\bm \gamma}_0 \\
&= \Sigma_{\Delta} - {\bm \gamma}_0^T  \Sigma_{\bm{W}}^{-1}  {\bm \gamma}_0
\end{align*}  
This shows that the asymptotic variance of $\widehat{\Delta}^\aug$ is no greater than that of  $\widehat{\Delta}$.
}}
\end{proof}

\blue{For completeness, we close by verifying that $\bm{\beta}_0$ minimizes the asymptotic variance of the augmented estimator.}

\begin{proof}
{\blue{For any $\bm{\beta} \in \mathbb{R}^p$, the asymptotic variance of the augmented estimator of 
$$\left\{ \widehat{\Delta} -  \bm{\beta}^T(\overline{\bm{W}}_1 - \overline{\bm{W}}_2 ) \right\}$$ 
is given by
\begin{align*}
\left(\dfrac{n_1 n_2}{n}\right) \left[ \text{Var}(\widehat{\Delta} - \Delta ) + \bm{\beta}^T \left\{ \text{Var}(\overline{\bm{W}}_1 -\overline{\bm{W}}_2)\right\} \bm{\beta} - 2\bm{\beta}^T \left\{ \text{Cov}(\overline{\bm{W}}_1 - \overline{\bm{W}}_2), \widehat{\Delta} - \Delta\right\}\right].
\end{align*}
The minimizer of this expression with respect to $\bm{\beta}$ is therefore 
$$ \left[ \text{Var} \left\{ \left(\dfrac{n_1 n_2}{n}\right)^{1/2}(\overline{\bm{W}}_1 - \overline{\bm{W}}_2) \right\}\right]^{-1} \text{Cov} \left\{ \left(\dfrac{n_1 n_2}{n}\right)^{1/2} (\overline{\bm W}_1 - \overline{\bm W}_2),  \left(\dfrac{n_1 n_2}{n}\right)^{1/2}(\widehat{\Delta} - \Delta) \right\}.$$ 
Note that
\begin{align*}
  \text{Var} \left\{ \left(\dfrac{n_1 n_2}{n}\right)^{1/2}(\overline{\bm{W}}_1 - \overline{\bm{W}}_2) \right\} = \rho_2 \text{Var}(\bm{W}_{i1}) +  \rho_1 \text{Var}(\bm{W}_{i2}) +  o(1) = \Sigma_{\bm{W}} + o(1) 
\end{align*}
and
\begin{align*}
  \text{Cov} \left\{ \left(\dfrac{n_1 n_2}{n}\right)^{1/2} (\overline{\bm W}_1 - \overline{\bm W}_2),  \left(\dfrac{n_1 n_2}{n}\right)^{1/2}(\widehat{\Delta} - \Delta) \right\} &= \rho_2  \text{Cov}(\bm{W}_{i1}, \Psi_{i1}) +  \rho_1  \text{Cov}(\bm{W}_{i2}, \Psi_{i2}) + o(1)\\
  &= {\bm \gamma} _0 + o(1). 
\end{align*}
It then follows that $\bm{\beta}_0 = \Sigma_{\bm{W}}^{-1} {\bm \gamma} _0$ minimizes the asymptotic variance of the augmented estimator.
}}
\end{proof}

\newpage
\section{Supplemental Simulation Results}
\subsection{Main Analysis}
\subsubsection{Additional results for \texorpdfstring{$n_1 = n_2 = 200$}{n1 = n2 = 200}}

\begin{figure}[ht]
    \centering
    \includegraphics[width=\textwidth]{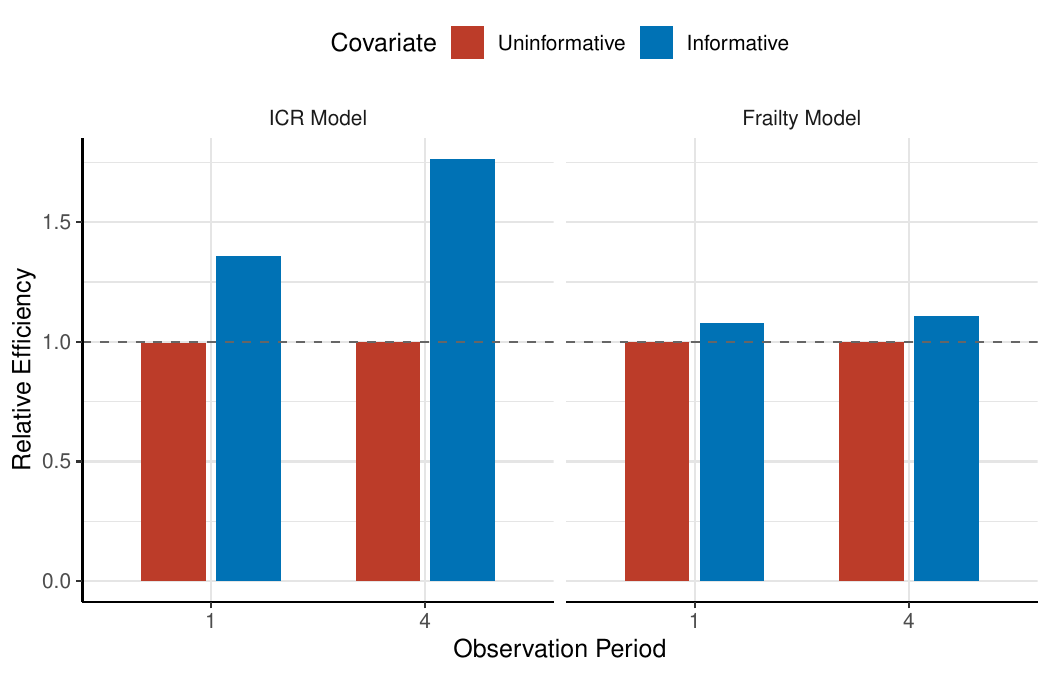}
    \caption{{\blue{Relative efficiency of the covariate adjusted AUMCF estimator with an informative covariate (blue) and an uninformative  covariate (red)  with treatment arm sample sizes of $n_1 = n_2 = 200$  under the independent competing risk (ICR) and frailty models for observation periods of $\tau = 1, 4$. The relative efficiency is defined as the ratio of the mean squared error of the unadjusted estimator relative to the adjusted estimator.}}}
    \label{fig:covariate_adjustment}
\end{figure}

\clearpage

\subsubsection{Results for \texorpdfstring{$n_1 = n_2 = 100$}{n1 = n2 = 100}}

\begin{figure}[ht]
    \centering
    \includegraphics[width=\textwidth]{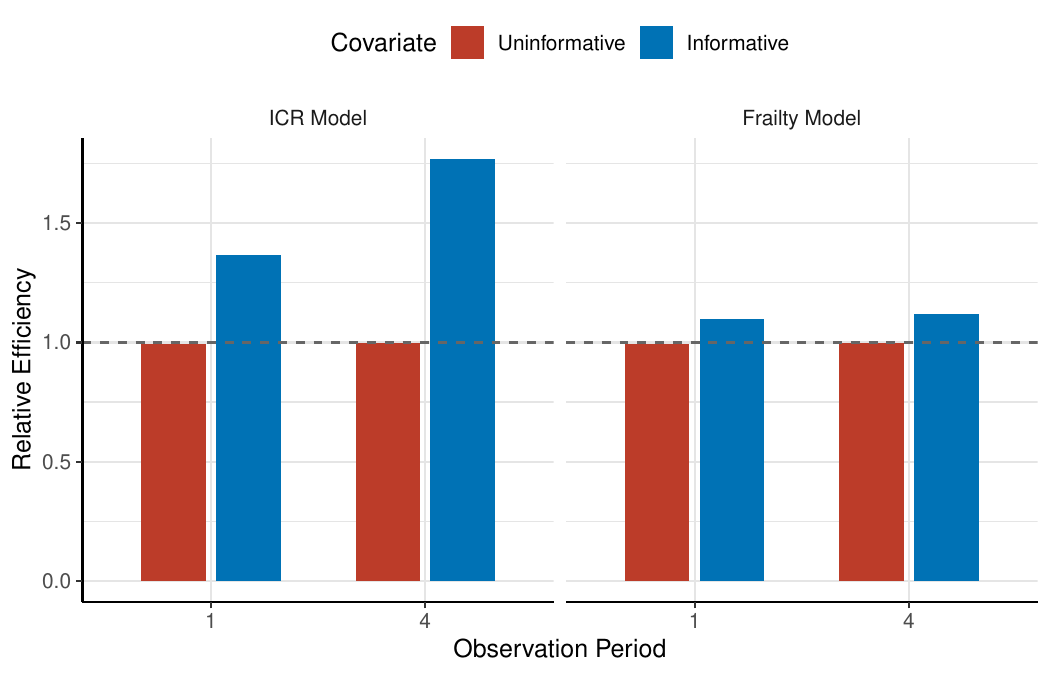}
    \caption{{\blue{Relative efficiency of the covariate adjusted AUMCF estimator with an informative covariate (blue) and an uninformative  covariate (red)  with treatment arm sample sizes of $n_1 = n_2 = 100$  under the independent competing risk and frailty models for observation periods of $\tau = 1, 4$.}}}
    \label{fig:covariate_100}
\end{figure}

\begin{table}[htbp]
\centering

\textbf{A: Null Setting} \\

\vspace{0.5em}

\scalebox{0.7}{
\begin{tabular}{lrrrr|rrrr}
\toprule
& \multicolumn{4}{c|}{$\tau = 1$} & \multicolumn{4}{c}{$\tau = 4$} \\
\cmidrule(lr){2-5} \cmidrule(lr){6-9}
\textbf{ICR Model} & Bias & ESE & ASE & Type I Error & Bias & ESE & ASE & Type I Error \\
\midrule
AUMCF         & -0.0008 & 0.084 & 0.082 & 0.058 & 0.0060 & 0.761 & 0.741 & 0.061 \\
CoxPH        & 0.0154  & 0.181 & 0.177 & 0.050 & 0.0113 & 0.159 & 0.155 & 0.049 \\
Frailty       & 0.0209  & 0.165 & 0.161 & 0.048 & 0.0093 & 0.103 & 0.103 & 0.046 \\
LWYY          & 0.0111  & 0.163 & 0.156 & 0.054 & 0.0058 & 0.102 & 0.100 & 0.055 \\
NB            & 0.0111  & 0.162 & 0.160 & 0.047 & 0.0057 & 0.102 & 0.103 & 0.048 \\
LWR      & 0.0173  & 0.186 & 0.187 & 0.045 & 0.0134 & 0.167 & 0.166 & 0.048 \\
SWR     & 0.0169  & 0.187 & 0.189 & 0.045 & 0.0149 & 0.170 & 0.169 & 0.048 \\
\midrule
\textbf{Frailty Model} & \multicolumn{4}{c|}{} & \multicolumn{4}{c}{} \\
AUMCF         & -0.0003 & 0.122 & 0.119 & 0.050 & -0.0054 & 1.115 & 1.086 & 0.055 \\
CoxPH        & 0.0305  & 0.246 & 0.233 & 0.049 & 0.0194 & 2.068 & 2.000 & 0.051 \\
Frailty       & 0.1073  & 0.400 & 0.333 & 0.063 & 0.0926 & 0.376 & 0.326 & 0.059 \\
LWYY          & 0.0473  & 0.326 & 0.291 & 0.060 & 0.0372 & 0.290 & 0.263 & 0.062 \\
NB            & 0.0543  & 0.395 & 0.311 & 0.063 & 0.0484 & 0.341 & 0.289 & 0.075 \\
LWR      & 0.0264  & 0.249 & 0.243 & 0.048 & 0.0225 & 0.215 & 0.212 & 0.050 \\
SWR     & 0.0265  & 0.250 & 0.243 & 0.049 & 0.0226 & 0.216 & 0.213 & 0.049 \\
\bottomrule
\end{tabular}}
\vspace{1em}

\textbf{B: Non-null Setting} \\

\vspace{0.5em}

\scalebox{0.7}{
\begin{tabular}{lrrrrr|rrrrr}
\toprule
& \multicolumn{5}{c|}{$\tau = 1$} & \multicolumn{5}{c}{$\tau = 4$} \\
\cmidrule(lr){2-6} \cmidrule(lr){7-11}
\textbf{ICR Model} & Bias & ESE & ASE & Power & CP & Bias & ESE & ASE & Power & CP \\
\midrule
AUMCF         & -0.0005 & 0.093 & 0.090 & 0.543 & 0.944 & 0.0479 & 0.841 & 0.834 & 0.823 & 0.951 \\
CoxPH        & 0.0099  & 0.127 & 0.171 & 0.376 & 0.957 & 0.0055 & 0.123 & 0.154 & 0.458 & 0.932 \\
Frailty       & 0.0094  & 0.108 & 0.148 & 0.586 & 0.955 & 0.0009 & 0.071 & 0.095 & 0.895 & 0.955 \\
LWYY          & 0.0061  & 0.107 & 0.144 & 0.649 & 0.948 & 0.0059 & 0.070 & 0.092 & 0.939 & 0.939 \\
NB            & 0.0062  & 0.107 & 0.148 & 0.630 & 0.949 & 0.0059 & 0.070 & 0.095 & 0.931 & 0.954 \\
LWR      & 0.0259  & 0.237 & 0.243 & 0.327 & - & 0.0259 & 0.212 & 0.209 & 0.302 & - \\
SWR     & 0.0252  & 0.223 & 0.232 & 0.234 & - & 0.0266 & 0.199 & 0.197 & 0.164 & - \\
\midrule
\textbf{Frailty Model} & \multicolumn{5}{c|}{} & \multicolumn{5}{c}{} \\
AUMCF         & -0.0052 & 0.143 & 0.141 & 0.208 & 0.946 & 0.0449 & 1.328 & 1.288 & 0.255 & 0.947 \\
CoxPH        & 0.0150  & 0.202 & 0.227 & 0.102 & 0.952 & 0.0054 & 1.800 & 1.964 & 0.109 & 0.943 \\
Frailty       & 0.0584  & 0.261 & 0.323 & 0.183 & 0.925 & 0.0614 & 0.260 & 0.318 & 0.168 & 0.942 \\
LWYY          & 0.0216  & 0.221 & 0.285 & 0.240 & 0.937 & 0.0296 & 0.202 & 0.202 & 0.266 & 0.937 \\
NB            & 0.0274  & 0.232 & 0.297 & 0.250 & 0.928 & 0.0378 & 0.225 & 0.284 & 0.265 & 0.931 \\
LWR      & 0.0410  & 0.277 & 0.276 & 0.096 & - & 0.0326 & 0.243 & 0.235 & 0.081 & - \\
SWR     & 0.0392  & 0.275 & 0.273 & 0.094 & - & 0.0328 & 0.242 & 0.234 & 0.074 & - \\
\bottomrule
\end{tabular}}
\caption{{\blue{\footnotesize Operating characteristics of the AUMCF estimator and the comparison methods with treatment arm sample sizes of $n_1 = n_2 = 100$  under the independent competing risk and frailty models for observation periods of $\tau = 1, 4$.  The comparison methods include the Cox Proportional Hazards Model (CoxPH), the shared frailty model (Frailty), the LWYY method (LWYY), negative binomial regression (NB), the standard win ratio (SWR), and the last-event-assisted win ratio (LWR). ESE is the empirical standard error across simulation replications.  ASE is the mean of the asymptotic standard errors across replicates.  CP is the empirical coverage probability of the 95\% confidence interval.  Results are shown for (A) type I error in the null setting corresponding to no difference in the AUMCFs across treatment arms and (B) power in a non-null setting where the difference in the AUMCFs across treatment arms is non-zero.  }}}
\label{tab:c1ac2_n100}
\end{table}

\begin{table}[htbp]
\centering

\textbf{A: Null Setting} \\

\vspace{0.5em}

\scalebox{0.95}{
\begin{tabular}{lrrrr|rrrr}
\toprule
& \multicolumn{4}{c|}{$\tau = 4$} & \multicolumn{4}{c}{$\tau = 6$} \\
\cmidrule(lr){2-5} \cmidrule(lr){6-9}
Method & Bias & ESE & ASE & Type I Error & Bias & ESE & ASE & Type I Error \\
\midrule
AUMCF    & -0.0086 & 0.629 & 0.620 & 0.055 & -0.0185 & 1.145 & 1.129 & 0.054 \\
CoxPH   &  0.0109 & 0.160 & 0.157 & 0.051 &  0.0114 & 0.159 & 0.156 & 0.052 \\
Frailty  &  0.0107 & 0.122 & 0.123 & 0.047 &  0.0076 & 0.116 & 0.117 & 0.045 \\
LWYY     &  0.0062 & 0.122 & 0.119 & 0.053 &  0.0046 & 0.115 & 0.113 & 0.052 \\
NB       &  0.0065 & 0.125 & 0.124 & 0.049 &  0.0052 & 0.119 & 0.117 & 0.046 \\
LWR &  0.0150 & 0.167 & 0.168 & 0.047 &  0.0145 & 0.168 & 0.166 & 0.054 \\
SWR&  0.0152 & 0.171 & 0.170 & 0.047 &  0.0142 & 0.170 & 0.169 & 0.052 \\
\bottomrule
\end{tabular}}

\vspace{1em}

\textbf{B: Non-null Setting} \\

\vspace{0.5em}

\scalebox{0.95}{
\begin{tabular}{lrrrrr|rrrrr}
\toprule
& \multicolumn{5}{c|}{$\tau = 4$} & \multicolumn{5}{c}{$\tau = 6$} \\
\cmidrule(lr){2-6} \cmidrule(lr){7-11}
Method & Bias & ESE & ASE & Power & CP & Bias & ESE & ASE & Power & CP \\
\midrule
AUMCF    &  0.0038 & 0.692 & 0.684 & 0.598 & 0.943 &  0.0259 & 1.381 & 1.321 & 0.798 & 0.945 \\
CoxPH   &  0.0353 & 0.143 & 0.156 & 0.108 & 0.949 &  0.0442 & 0.142 & 0.156 & 0.106 & 0.940 \\
Frailty  &  0.0436 & 0.085 & 0.113 & 0.735 & 0.931 &  0.0455 & 0.079 & 0.108 & 0.821 & 0.915 \\
LWYY     &  0.0473 & 0.077 & 0.110 & 0.888 & 0.912 &  0.0579 & 0.074 & 0.105 & 0.945 & 0.863 \\
NB       &  0.0484 & 0.079 & 0.113 & 0.871 & 0.919 &  0.0595 & 0.075 & 0.107 & 0.935 & 0.864 \\
LWR &  0.0209 & 0.185 & 0.188 & 0.108 & - &  0.0178 & 0.183 & 0.184 & 0.091 & - \\
SWR&  0.0179 & 0.171 & 0.173 & 0.049 & - &  0.0147 & 0.170 & 0.171 & 0.042 & - \\
\bottomrule
\end{tabular}}
\caption{{\blue{Operating characteristics of the AUMCF estimator and the comparison methods with treatment arm sample sizes of $n_1 = n_2 = 100$  under the time-varying treatment effect model for observation periods of $\tau = 4, 6$.  The comparison methods include the Cox Proportional Hazards Model (CoxPH), the shared frailty model (Frailty), the LWYY method (LWYY), negative binomial regression (NB), the standard win ratio (SWR), and the last-event-assisted win ratio (LWR). ESE is the empirical standard error across simulation replications.  ASE is the mean of the asymptotic standard errors across replicates.  CP is the empirical coverage probability of the 95\% confidence interval.  Results are shown for (A) type I error in the null setting corresponding to no difference in the AUMCFs across treatment arms and (B) power in a non-null setting where the difference in the AUMCFs across treatment arms is non-zero.  }}}
\label{tab:c3_n100}
\end{table}

\clearpage
\subsubsection{Results for \texorpdfstring{$n_1 = n_2 = 400$}{n1 = n2 = 400}}
\begin{figure}[ht]
    \centering
    \includegraphics[width=\textwidth]{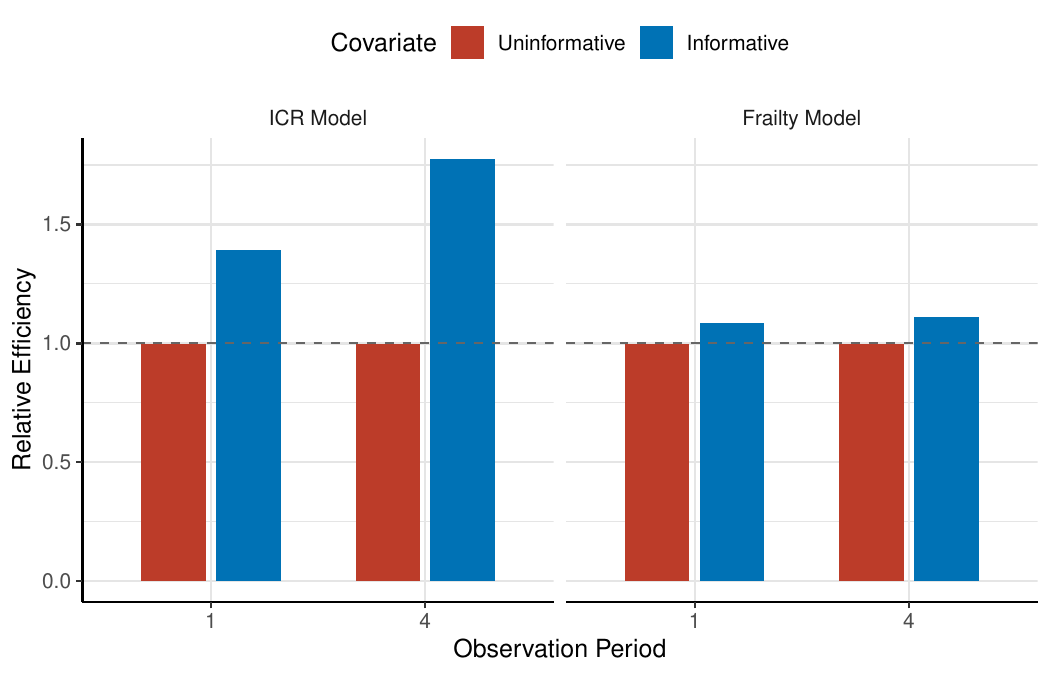}
    \caption{{\blue{Relative efficiency of the covariate adjusted AUMCF estimator with an informative covariate (blue) and an uninformative  covariate (red)  with treatment arm sample sizes of $n_1 = n_2 = 400$  under the independent competing risk and frailty models for observation periods of $\tau = 1, 4$. The relative efficiency is defined as the ratio of the mean squared error of the unadjusted estimator relative to the adjusted estimator.}}}
    \label{fig:covariate_400}
\end{figure}

\begin{table}[htbp]
\centering
\textbf{A: Null Setting} \\

\vspace{0.5em}

\scalebox{0.7}{
\begin{tabular}{lrrrr|rrrr}
\toprule
& \multicolumn{4}{c|}{$\tau = 1$} & \multicolumn{4}{c}{$\tau = 4$} \\
\cmidrule(lr){2-5} \cmidrule(lr){6-9}
\textbf{ICR Model} & Bias & ESE & ASE & Type I Error & Bias & ESE & ASE & Type I Error \\
\midrule
AUMCF         & 0.0000 & 0.042 & 0.041 & 0.053 & -0.0038 & 0.375 & 0.374 & 0.053 \\
CoxPH        & 0.0041 & 0.090 & 0.088 & 0.053 &  0.0022 & 0.077 & 0.077 & 0.049 \\
Frailty       & 0.0061 & 0.079 & 0.079 & 0.051 &  0.0014 & 0.050 & 0.051 & 0.049 \\
LWYY          & 0.0032 & 0.079 & 0.078 & 0.055 &  0.0007 & 0.050 & 0.050 & 0.054 \\
NB            & 0.0032 & 0.079 & 0.079 & 0.052 &  0.0007 & 0.050 & 0.051 & 0.050 \\
LWR      & 0.0036 & 0.093 & 0.092 & 0.052 &  0.0040 & 0.082 & 0.082 & 0.053 \\
SWR     & 0.0038 & 0.093 & 0.093 & 0.050 &  0.0041 & 0.083 & 0.083 & 0.051 \\
\midrule
\textbf{Frailty Model} & \multicolumn{4}{c|}{} & \multicolumn{4}{c}{} \\
AUMCF         & 0.0000 & 0.062 & 0.060 & 0.054 & -0.0019 & 0.554 & 0.549 & 0.048 \\
CoxPH        & 0.0061 & 0.117 & 0.116 & 0.049 &  0.0041 & 0.101 & 0.100 & 0.051 \\
Frailty       & 0.0269 & 0.177 & 0.167 & 0.057 &  0.0176 & 0.170 & 0.164 & 0.052 \\
LWYY          & 0.0118 & 0.153 & 0.148 & 0.056 &  0.0084 & 0.137 & 0.133 & 0.052 \\
NB            & 0.0135 & 0.177 & 0.153 & 0.062 &  0.0104 & 0.158 & 0.145 & 0.067 \\
LWR      & 0.0076 & 0.118 & 0.118 & 0.047 &  0.0069 & 0.105 & 0.104 & 0.052 \\
SWR     & 0.0078 & 0.118 & 0.118 & 0.050 &  0.0069 & 0.106 & 0.104 & 0.052 \\
\bottomrule
\end{tabular}}
\vspace{1em}

\textbf{B: Non-Null Setting} \\

\vspace{0.5em}

\scalebox{0.7}{
\begin{tabular}{lrrrrr|rrrrr}
\toprule
& \multicolumn{5}{c|}{$\tau = 1$} & \multicolumn{5}{c}{$\tau = 4$} \\
\cmidrule(lr){2-6} \cmidrule(lr){7-11}
\textbf{ICR Model} & Bias & ESE & ASE & Power & Coverage & Bias & ESE & ASE & Power & Coverage \\
\midrule
AUMCF         & -0.0015 & 0.046 & 0.045 & 0.991 & 0.949 & -0.0106 & 0.420 & 0.421 & 1.000 & 0.957 \\
CoxPH        & -0.0023 & 0.064 & 0.085 & 0.935 & 0.956 & -0.0033 & 0.058 & 0.077 & 0.967 & 0.941 \\
Frailty       & -0.0014 & 0.053 & 0.073 & 0.996 & 0.951 & -0.0048 & 0.034 & 0.047 & 1.000 & 0.951 \\
LWYY          & -0.0018 & 0.052 & 0.072 & 0.999 & 0.945 & -0.0010 & 0.032 & 0.046 & 1.000 & 0.961 \\
NB            & -0.0018 & 0.052 & 0.073 & 0.999 & 0.949 & -0.0010 & 0.032 & 0.047 & 1.000 & 0.962 \\
LWR      &  0.0135 & 0.123 & 0.120 & 0.879 & - &  0.0082 & 0.107 & 0.102 & 0.813 & - \\
SWR     &  0.0135 & 0.117 & 0.114 & 0.729 & - &  0.0087 & 0.101 & 0.096 & 0.468 & - \\
\midrule
\textbf{Frailty Model} & \multicolumn{5}{c|}{} & \multicolumn{5}{c}{} \\
AUMCF         &  0.0021 & 0.070 & 0.071 & 0.620 & 0.944 &  0.0103 & 0.666 & 0.652 & 0.746 & 0.940 \\
CoxPH        &  0.0051 & 0.103 & 0.113 & 0.284 & 0.933 & -0.0044 & 0.088 & 0.098 & 0.257 & 0.950 \\
Frailty       &  0.0201 & 0.124 & 0.162 & 0.539 & 0.939 &  0.0079 & 0.118 & 0.160 & 0.543 & 0.952 \\
LWYY          &  0.0115 & 0.109 & 0.144 & 0.646 & 0.944 &  0.0082 & 0.098 & 0.131 & 0.719 & 0.935 \\
NB            &  0.0115 & 0.113 & 0.149 & 0.654 & 0.937 &  0.0108 & 0.107 & 0.142 & 0.735 & 0.938 \\
LWR      &  0.0065 & 0.139 & 0.133 & 0.240 & - &  0.0050 & 0.116 & 0.114 & 0.195 & - \\
SWR     &  0.0062 & 0.137 & 0.132 & 0.225 & - &  0.0039 & 0.115 & 0.113 & 0.171 & - \\
\bottomrule
\end{tabular}}
\caption{{\blue{\footnotesize Operating characteristics of the AUMCF estimator and the comparison methods with treatment arm sample sizes of $n_1 = n_2 = 400$  under the independent competing risk and frailty models for observation periods of $\tau = 1, 4$.  The comparison methods include the Cox Proportional Hazards Model (CoxPH), the shared frailty model (Frailty), the LWYY method (LWYY), negative binomial regression (NB), the standard win ratio (SWR), and the last-event-assisted win ratio (LWR). ESE is the empirical standard error across simulation replications.  ASE is the mean of the asymptotic standard errors across replicates.  CP is the empirical coverage probability of the 95\% confidence interval.  Results are shown for (A) type I error in the null setting corresponding to no difference in the AUMCFs across treatment arms and (B) power in a non-null setting where the difference in the AUMCFs across treatment arms is non-zero.  }}}
\label{tab:c1ac2_n400}
\end{table}

\begin{table}[htbp]
\centering
\textbf{A: Null Setting} \\

\vspace{0.5em}

\scalebox{0.95}{
\begin{tabular}{lrrrr|rrrr}
\toprule
& \multicolumn{4}{c|}{$\tau = 4$} & \multicolumn{4}{c}{$\tau = 6$} \\
\cmidrule(lr){2-5} \cmidrule(lr){6-9}
Method & Bias & ESE & ASE & Type I Error & Bias & ESE & ASE & Type I Error \\
\midrule
AUMCF    & -0.0022 & 0.306 & 0.312 & 0.051 & 0.0233 & 0.557 & 0.569 & 0.052 \\
CoxPH   & 0.0019 & 0.076 & 0.078 & 0.033 & 0.0029 & 0.075 & 0.078 & 0.041 \\
Frailty  & 0.0022 & 0.057 & 0.061 & 0.034 & 0.0025 & 0.056 & 0.058 & 0.042 \\
LWYY     & 0.0015 & 0.058 & 0.060 & 0.050 & 0.0020 & 0.055 & 0.057 & 0.040 \\
NB       & 0.0019 & 0.060 & 0.061 & 0.044 & 0.0024 & 0.057 & 0.059 & 0.045 \\
LWR & 0.0027 & 0.079 & 0.082 & 0.045 & 0.0070 & 0.080 & 0.082 & 0.044 \\
SWR& 0.0024 & 0.081 & 0.084 & 0.043 & 0.0073 & 0.080 & 0.083 & 0.042 \\
\bottomrule
\end{tabular}}

\vspace{1em}

\textbf{B: Non-Null Setting} \\

\vspace{0.5em}

\scalebox{0.95}{
\begin{tabular}{lrrrrr|rrrrr}
\toprule
& \multicolumn{5}{c|}{$\tau = 4$} & \multicolumn{5}{c}{$\tau = 6$} \\
\cmidrule(lr){2-6} \cmidrule(lr){7-11}
Method & Bias & ESE & ASE & Power & CP & Bias & ESE & ASE & Power & CP \\
\midrule
AUMCF    & 0.0162 & 0.336 & 0.344 & 0.993 & 0.954 & 0.0288 & 0.667 & 0.668 & 1.000 & 0.947 \\
CoxPH   & 0.0303 & 0.066 & 0.078 & 0.334 & 0.936 & 0.0342 & 0.067 & 0.078 & 0.315 & 0.940 \\
Frailty  & 0.0370 & 0.042 & 0.056 & 1.000 & 0.861 & 0.0400 & 0.040 & 0.053 & 0.999 & 0.826 \\
LWYY     & 0.0445 & 0.039 & 0.055 & 1.000 & 0.808 & 0.0543 & 0.036 & 0.053 & 1.000 & 0.648 \\
NB       & 0.0453 & 0.040 & 0.056 & 1.000 & 0.806 & 0.0555 & 0.036 & 0.053 & 1.000 & 0.649 \\
LWR & 0.0013 & 0.091 & 0.092 & 0.261 & - & 0.0019 & 0.087 & 0.090 & 0.224 & - \\
SWR& 0.0006 & 0.083 & 0.085 & 0.053 & - & 0.0012 & 0.080 & 0.084 & 0.045 & - \\
\bottomrule
\end{tabular}}
\caption{{\blue{Operating characteristics of the AUMCF estimator and the comparison methods with treatment arm sample sizes of $n_1 = n_2 = 400$  under the time-varying treatment effect model for observation periods of $\tau = 4, 6$.  The comparison methods include the Cox Proportional Hazards Model (CoxPH), the shared frailty model (Frailty), the LWYY method (LWYY), negative binomial regression (NB), the standard win ratio (SWR), and the last-event-assisted win ratio (LWR). ESE is the empirical standard error across simulation replications.  ASE is the mean of the asymptotic standard errors across replicates.  CP is the empirical coverage probability of the 95\% confidence interval.  Results are shown for (A) type I error in the null setting corresponding to no difference in the AUMCFs across treatment arms and (B) power in a non-null setting where the difference in the AUMCFs across treatment arms is non-zero.  }}}
\label{tab:c3_n400}
\end{table}

\clearpage
\subsection{Sensitivity Analysis}
{\blue{Here we evaluate the impact of survival bias on the AUMCF and the comparison methods by considering scenarios 1 and 2 with $n_1 = n_2 = 200$. All parameters are the same as in the main text for the null setting with the exception of the terminal event rate in the first group, that is, $\lambda_{D_2} =0.2$ and $\lambda_{D_1} \in (0.10, 0.15, 0.18)$. Results are summarized across 500 simulation replicates.\\ }} 

{\blue{As expected, the impact of survival bias is the most severe for all methods for $\tau = 4$ relative to $\tau = 1$. In scenario 1 under the independent competing risk model, negative binomial regression and the LWYY method are the only parametric methods that are minimally impacted. The AUMCF is most impacted when $\lambda_{D_1} = 0.10$ and $\tau = 4$, but exhibits relatively less bias than the last-event assisted win ratio (LWR) and the standard win ratio (SWR). In scenario 2 under the frailty model, the Cox proportional hazards model (CoxPH) and the shared frailty model (Frailty) are the only parametric methods that are not impacted by survival bias. The AUMCF is again substantially impacted when $\lambda_{D_1} = 0.10$ and $\tau = 4$ and has slightly higher bias than the win ratio methods (relative to their respective ESEs).}}

\begin{table}[H]
\centering
\begin{subtable}[t]{0.48\textwidth}
\centering
\caption{$\tau = 1$}
\begin{tabular}{|lrrr|}
\toprule
Method & Est. & ASE & ESE\\
\hline
$\lambda_{D_1} = 0.18$ &  &  &  \\
\hline 
AUMCF & 0.003 & 0.058 & 0.059\\
CoxPH & 0.987 & 0.125 & 0.130\\
Frailty & 1.003 & 0.112 & 0.116\\
LWYY & 1.005 & 0.110 & 0.116\\
NB & 1.005 & 0.112 & 0.116\\
LWR & 1.047 & 0.136 & 0.139\\
SWR & 1.048 & 0.137 & 0.141\\
\hline
$\lambda_{D_1} = 0.15$ &  &  &  \\
\hline 
AUMCF & 0.008 & 0.058 & 0.059\\
CoxPH & 0.961 & 0.125 & 0.125\\
Frailty & 0.993 & 0.112 & 0.113\\
LWYY & 1.006 & 0.109 & 0.115\\
NB & 1.006 & 0.111 & 0.115\\
LWR & 1.105 & 0.144 & 0.147\\
SWR & 1.106 & 0.145 & 0.148\\
\hline
$\lambda_{D_1} = 0.10$ &  &  &  \\
\hline 
AUMCF & 0.016 & 0.058 & 0.058\\
CoxPH & 0.920 & 0.126 & 0.120\\
Frailty & 0.974 & 0.111 & 0.110\\
LWYY & 1.006 & 0.109 & 0.113\\
NB & 1.006 & 0.111 & 0.113\\
LWR & 1.212 & 0.159 & 0.163\\
SWR & 1.214 & 0.161 & 0.167\\
\bottomrule
\end{tabular}
\end{subtable}
\hfill
\begin{subtable}[t]{0.48\textwidth}
\centering
\caption{$\tau = 4$}
\begin{tabular}{|lrrr|}
\toprule
Method & Est. & ASE & ESE\\
\hline
$\lambda_{D_1} = 0.18$ &  &  &  \\
\hline
AUMCF & 0.136 & 0.528 & 0.529\\
CoxPH & 0.995 & 0.109 & 0.106\\
Frailty & 0.985 & 0.072 & 0.069\\
LWYY & 0.997 & 0.070 & 0.070\\
NB & 0.997 & 0.071 & 0.070\\
LWR & 1.068 & 0.123 & 0.118\\
SWR & 1.064 & 0.125 & 0.120\\
\hline
$\lambda_{D_1} = 0.15$ &  &  &  \\
\hline
AUMCF & 0.355 & 0.529 & 0.525\\
CoxPH & 0.967 & 0.109 & 0.102\\
Frailty & 0.965 & 0.071 & 0.066\\
LWYY & 0.997 & 0.069 & 0.067\\
NB & 0.997 & 0.071 & 0.067\\
LWR & 1.170 & 0.135 & 0.128\\
SWR & 1.165 & 0.137 & 0.132\\
\hline
$\lambda_{D_1} = 0.10$ &  &  &  \\
\hline
AUMCF & 0.806 & 0.528 & 0.540\\
CoxPH & 0.926 & 0.109 & 0.099\\
Frailty & 0.935 & 0.070 & 0.063\\
LWYY & 1.000 & 0.068 & 0.068\\
NB & 1.000 & 0.069 & 0.068\\
LWR & 1.390 & 0.161 & 0.156\\
SWR & 1.386 & 0.166 & 0.165\\
\bottomrule
\end{tabular}
\end{subtable}
\caption{{\blue{Performance of the various methods for Scenario 1 (independent competing risk Model) under the null with $\lambda_{D_2} = 0.2$ and $\lambda_{D_1} \in \{0.18, 0.15, 0.10\}$ based on the point estimate (Est.), asymptotic standard error (ASE) and empirical standard error (ESE). The difference in AUMCFs across treatment arms is compared to the Cox proportional hazards model (CoxPH), the LWYY method (LWYY), negative binomial regression (NB), the shared frailty model (Frailty), the last-event-assisted win ratio (LWR), and the standard win ratio (SWR).}}}
\end{table}

\begin{table}[H]
\centering
\begin{subtable}[t]{0.48\textwidth}
\centering
\caption{$\tau = 1$}
\begin{tabular}{|lrrr|}
\toprule
Method & Est. & ASE & ESE \\
\hline
$\lambda_{D_1} = 0.18$ &  &  &  \\
\hline 
AUMCF & 0.007 & 0.085 & 0.088\\
CoxPH & 1.008 & 0.165 & 0.173\\
Frailty & 1.048 & 0.233 & 0.251\\
LWYY & 1.042 & 0.207 & 0.224\\
NB & 1.034 & 0.215 & 0.234\\
LWR & 1.032 & 0.172 & 0.178\\
SWR & 1.034 & 0.173 & 0.179\\
\hline
$\lambda_{D_1} = 0.15$ &  &  &  \\
\hline 
AUMCF & 0.016 & 0.086 & 0.089\\
CoxPH & 0.992 & 0.165 & 0.171\\
Frailty & 1.047 & 0.232 & 0.257\\
LWYY & 1.068 & 0.207 & 0.234\\
NB & 1.049 & 0.214 & 0.240\\
LWR & 1.063 & 0.178 & 0.179\\
SWR & 1.066 & 0.179 & 0.179\\
\hline
$\lambda_{D_1} = 0.10$ &  &  &  \\
\hline 
AUMCF & 0.041 & 0.089 & 0.089\\
CoxPH & 0.970 & 0.166 & 0.169\\
Frailty & 1.055 & 0.228 & 0.247\\
LWYY & 1.136 & 0.207 & 0.240\\
NB & 1.090 & 0.213 & 0.239\\
LWR & 1.119 & 0.188 & 0.188\\
SWR & 1.128 & 0.190 & 0.190\\
\bottomrule
\end{tabular}
\end{subtable}
\hfill
\begin{subtable}[t]{0.48\textwidth}
\centering
\caption{$\tau = 4$}
\begin{tabular}{|lrrr|}
\toprule
Method & Est. & ASE & ESE \\
\hline
$\lambda_{D_1} = 0.18$ &  &  &  \\
\hline 
AUMCF & 0.186 & 0.786 & 0.754\\
CoxPH & 1.004 & 0.141 & 0.139\\
Frailty & 1.037 & 0.231 & 0.236\\
LWYY & 1.057 & 0.188 & 0.192\\
NB & 1.034 & 0.205 & 0.216\\
LWR & 1.032 & 0.151 & 0.155\\
SWR & 1.034 & 0.152 & 0.155\\
\hline
$\lambda_{D_1} = 0.15$ &  &  &  \\
\hline 
AUMCF & 0.569 & 0.824 & 0.830\\
CoxPH & 0.999 & 0.142 & 0.144\\
Frailty & 1.042 & 0.228 & 0.235\\
LWYY & 1.137 & 0.188 & 0.217\\
NB & 1.070 & 0.204 & 0.227\\
LWR & 1.068 & 0.157 & 0.164\\
SWR & 1.073 & 0.158 & 0.166\\
\hline
$\lambda_{D_1} = 0.10$ &  &  &  \\
\hline 
AUMCF & 1.255 & 0.895 & 0.886\\
CoxPH & 0.980 & 0.142 & 0.137\\
Frailty & 1.030 & 0.223 & 0.224\\
LWYY & 1.276 & 0.188 & 0.237\\
NB & 1.116 & 0.201 & 0.230\\
LWR & 1.141 & 0.168 & 0.170\\
SWR & 1.152 & 0.170 & 0.172\\
\bottomrule
\end{tabular}
\end{subtable}
\caption{{\blue{Performance of the various methods for Scenario 2 (Frailty Model) under the null with $\lambda_{D_2} = 0.2$ and $\lambda_{D_1} \in \{0.18, 0.15, 0.10\}$ based on the point estimate (Est.), asymptotic standard error (ASE), and empirical standard error (ESE).  The difference in AUMCFs across treatment arms is compared to the Cox proportional hazards model (CoxPH), the LWYY method (LWYY), negative binomial regression (NB), the shared frailty model (Frailty), the last-event-assisted win ratio (LWR), and the standard win ratio (SWR).}}}
\end{table}

\section{Supplemental Data Analysis}
To further explore the association between bucindolol and the disease-burden endpoint, we conducted various sensitivity analyses. For reference, in the primary analysis the 48-month AUMCFs were 39.7 months with bucindolol and 43.5 months with placebo, for a difference of 3.73 months (95\% CI, 1.32 to 6.15 months, $P = 0.0025$). If we remove death as an event of interest, instead regarding it purely as a competing risk, and compare the disease burdens attributable to the remaining, non-fatal events (hospitalization, myocardial infarction, and heart transplantation), the 48-month AUMCFs are 27.9 months for bucindolol and 30.6 months for placebo. The difference is 2.69 months (95\% CI, 0.87 to 4.51 months, $P = 0.0037$) in favor of bucindolol. Conversely, if we retain death among the events of interest and increase its weight to double that of a non-fatal event, the 48-month AUMCFs are 51.6 months for bucindolol and 56.3 months for placebo. The difference is 4.78 months (95\% CI, 1.32 to 8.23 months, $P = 0.0067$) in favor of bucindolol. Together, these analyses suggest that while death may contribute to the overall reduction in disease burden with bucindolol, the primary driver of the treatment difference is a reduction in the burden of non-fatal events.

\section{Supplemental Figures}

\begin{figure}[ht]
    \centering
    \includegraphics[width=\textwidth]{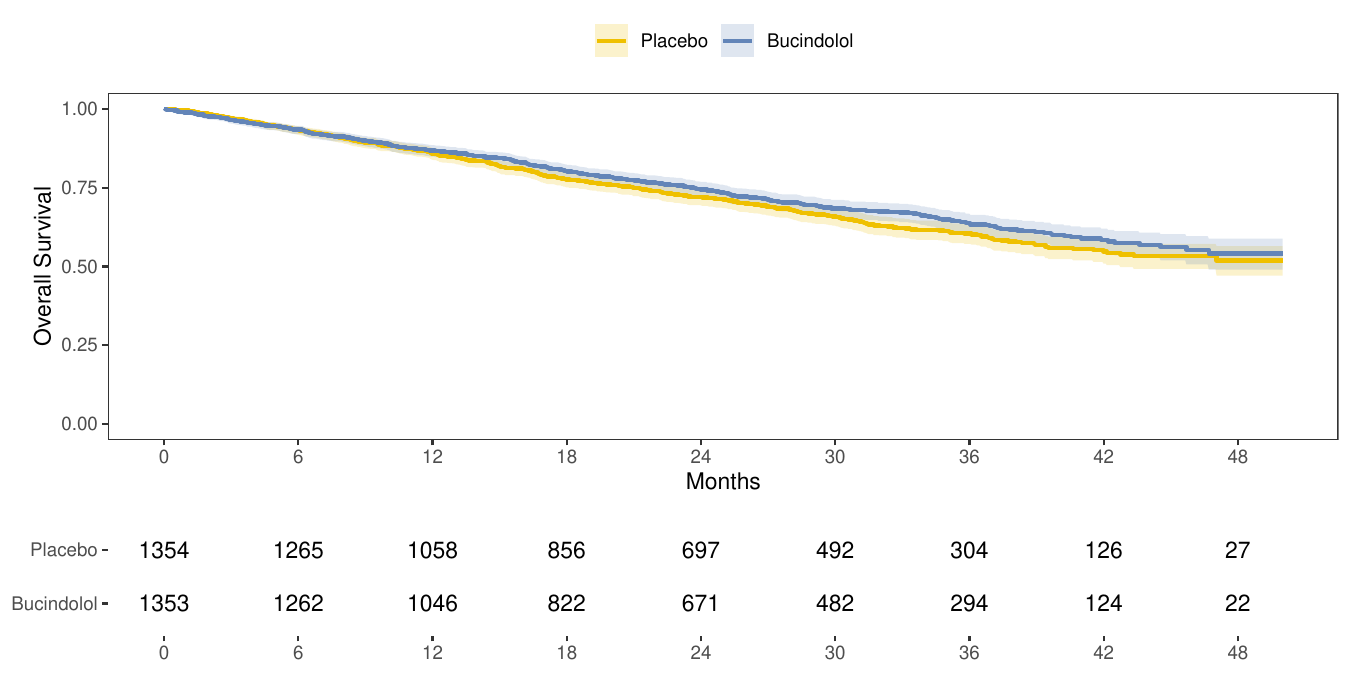}
    \caption{\textbf{Kaplan-Meier curves for overall survival based on primary data from the Beta-Blocker Evaluation of Survival (BEST) trial.}}
    \label{supp:fig:km-os}
\end{figure}

\begin{figure}[ht]
    \centering
    \includegraphics[width=\textwidth]{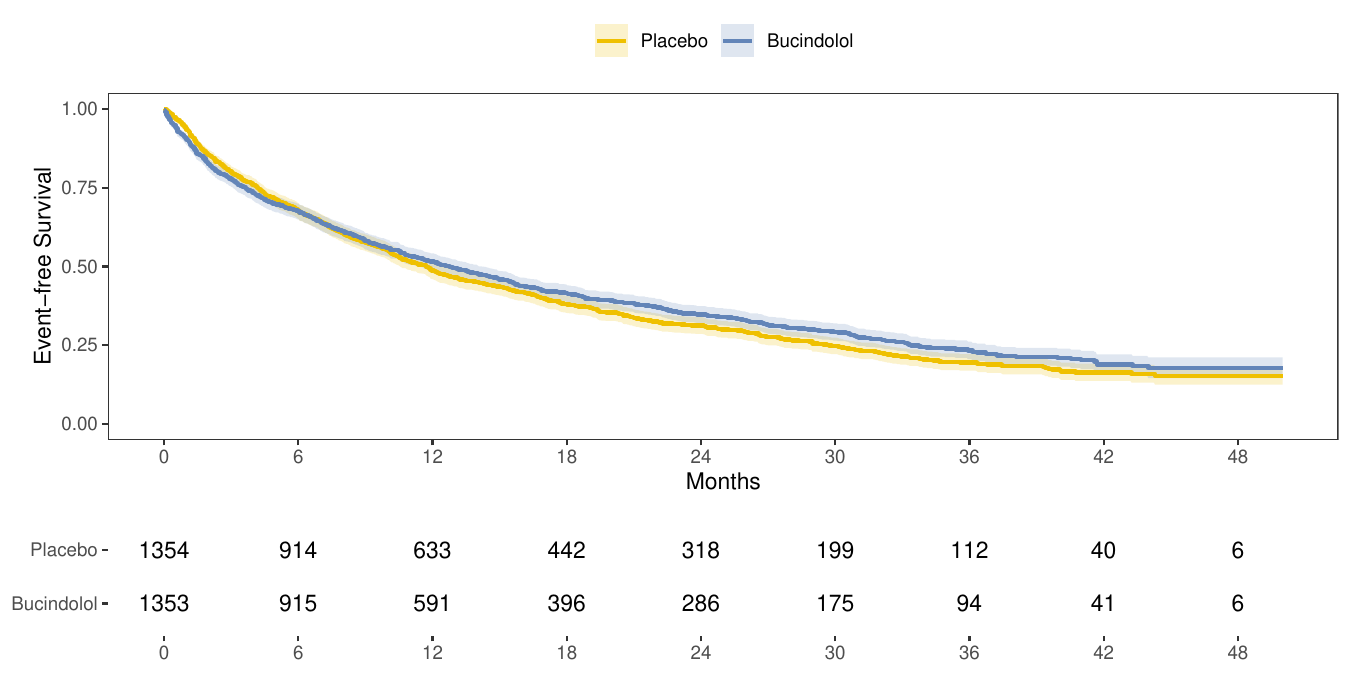}
    \caption{\textbf{Kaplan-Meier curves for event-free survival based on primary data from the BEST trial.} The composite disease-burden endpoint was defined as the time to hospitalization from any cause, myocardial infarction, heart transplantation, and death, whichever occurred first.}
    \label{supp:fig:km-composite}
\end{figure}

\clearpage
\bibliographystyle{apalike}
\bibliography{biomsample}